\documentclass[12pt]{article}
\pdfoutput=1
\usepackage{graphicx}
\usepackage{epstopdf}
\usepackage[body={17.5cm, 21cm},right=2cm]{geometry}
\usepackage{amssymb}
\usepackage{amsmath}

\newcommand\eea{\end{eqnarray}}
\newcommand\bea{\begin{eqnarray}}

\newcommand\mpl{M_{\rm Pl}}

\def\beq{\begin{equation}}
\def\eeq{\end{equation}}
\def\d{\partial}

\def\d{\partial}

\newcommand{\be}{\begin{equation}}
\newcommand{\ee}{\end{equation}}
\newcommand{\ba}{\begin{align}}
\newcommand{\ea}{\end{align}}
\newcommand{\bg}{\begin{gather}}
\newcommand{\eg}{\end{gather}}
\newcommand{\bseq}{\begin{subequations}}
\newcommand{\eseq}{\end{subequations}}

\begin{document}

\vspace{5mm}
\vspace{0.5cm}
\begin{center}

\def\thefootnote{\fnsymbol{footnote}}

{\Large \bf On Loops in Inflation III:\\[0.4cm]
Time Independence of $\zeta$ in Single Clock Inflation}
\\[0.5cm]

{\large  Guilherme L.~Pimentel$^1$, Leonardo Senatore$^{2,3}$, and Matias Zaldarriaga$^4$}
\\[0.5cm]

{\normalsize { \sl $^{1}$  Joseph Henry Laboratories, Princeton University, Princeton, NJ 08544}}\\
\vspace{.3cm}

{\normalsize { \sl $^{2}$ Stanford Institute for Theoretical Physics, Stanford University, Stanford, CA 94306}}\\
\vspace{.3cm}

{\normalsize { \sl $^{3}$ Kavli Institute for Particle Astrophysics and Cosmology, Stanford University and SLAC,\\ Menlo Park, CA 94025}}\\
\vspace{.3cm}

{\small \normalsize{\sl $^{4}$ Institute for Advanced Study, Einstein Drive,  Princeton, NJ 08540}}

\end{center}

%\vspace{.8cm}

%\hrule \vspace{0.3cm}
%{\small  \noindent \textbf{Abstract} \\[0.3cm]
\noindent

%\vspace{0.5cm}  \hrule
\def\thefootnote{\arabic{footnote}}
\setcounter{footnote}{0}

\vspace{.8cm}
\hrule \vspace{0.3cm}
{\small  \noindent \textbf{Abstract} \\[0.3cm]
\noindent Studying loop corrections to inflationary perturbations, with particular emphasis on infrared factors, is important to understand the consistency of the inflationary theory, its predictivity and to establish the existence of the slow-roll eternal inflation phenomena and its recently found volume bound. In this paper we prove that the $\zeta$ correlation function is time-independent at one-loop level in single clock inflation. While many of the one-loop diagrams lead to a time-dependence when considered individually, the time-dependence beautifully cancels out in the overall sum. We identify two subsets of diagrams that cancel separately due to different physical reasons. The first cancellation is related to the change of the background cosmology due to the renormalization of the stress tensor. It results in  a cancellation between the non-1PI diagrams and some of the diagrams made with quartic vertices. The second subset of diagrams that cancel is made up of cubic operators, plus the remaining quartic ones. We are able to write the sum of these diagrams as the integral over a specific three-point function between two very short wavelengths and one very long one. We then apply the consistency condition for this three-point function in the squeezed limit to show that the sum of these diagrams cannot give rise to a time dependence. This second cancellation is thus a consequence of the fact that in single clock inflation the attractor nature of the solution implies that a long wavelength $\zeta$ perturbation is indistinguishable from a trivial rescaling of the background, and so results in no physical effect on short wavelength modes.  
%This exhaust all the one loop diagrams, proving that the $\zeta$ correlator is time independent at one loop~level. 
}
 \vspace{0.3cm}
\hrule

\section{Introduction}

\subsection{Motivation}
The purpose of this paper is to prove that in single clock inflation, where there is only one relevant degree of freedom during inflation, the correlation function of the curvature perturbation $\zeta$ for separations outside the horizon is time independent at one loop level. We believe this to be a very important result to prove for several reasons. As it becomes more and more likely that Inflation was part of the early history of our Universe it becomes more and more important to understand how the theory behaves at quantum level, even if the expected corrections are small. We could make an analogy with the 1950s when QED was studied to all orders in perturbation theory. Similarly to what happened in that case, it is not so obvious that quantum corrections are as small as one might expect. While a simple parametric analysis tells that the corrections to the curvature perturbation should be of order
\be
\langle\zeta^2\rangle_{\rm 1-loop}\sim \langle\zeta^2\rangle_{tree}^2\sim 10^{-9}\langle\zeta^2\rangle_{tree}\ ,
\ee
no symmetry forbids the presence of potentially large infrared factors, such as 
\be
\langle\zeta_k^2\rangle_{\rm 1-loop}\sim k^3 \langle\zeta_k^2\rangle_{tree}^2 \log(k L)\ ,
\ee
where $L$ is the comoving size of the inflationary space, or of the form
\be
\langle\zeta_k^2\rangle_{\rm 1-loop}\sim k^3 \langle\zeta_k^2\rangle_{tree}^2  H t\ ,
\ee
where $H$ is the Hubble constant during inflation and $t$ is time. All these terms have appeared in partial calculations of the one-loop corrections to the power spectrum~\cite{Weinberg:2005vy}. 

{\bf $Log(H/\mu)$ effects:} Additionally, infrared effects of the form
\be
\langle\zeta_k^2\rangle_{\rm 1-loop}\sim k^3 \langle\zeta_k^2\rangle_{tree}^2 \log(k/\mu)\ ,
\ee
with $\mu$ being the renormalization scale of the theory, have been found in several papers (see references in~\cite{Senatore:2009cf}). Strictly speaking, a correction of the form $\log(k/\mu)$ is not allowed by symmetries, representing a breaking of zero-mode gauge invariance $x\rightarrow \lambda x, a\rightarrow a/\lambda$, where $a$ is the scale factor of the FRW metric. Its presence was due to a mistake in the implementation of a diff. invariant regularization, and some of us addressed this issue in~\cite{Senatore:2009cf}, where it was shown that the logarithmic running takes the form
\be
\langle\zeta_k^2\rangle_{\rm 1-loop}\sim k^3 \langle\zeta_k^2\rangle^2 \log(H/\mu)\ .
\ee
%This result is very intuitive, the logarithmic running is small if the renormalization scale is chosen to be close to $H$, the energy scale of the inflationary fluctuations, as it is clearly seen from the Effective Field Theory of Inflation~\cite{Cheung:2007st}. 
Notice that if a result of the form $\log(k/\mu)$ were to be correct, then the effect could have been potentially very large when $k\rightarrow0$. 
%Indeed, the choice of the natural scale $E$ that appears in the $\log(E/\mu)$ to make the correction small is an infrared question, and so potentially large.

Contrary to the case of $\log(k/\mu)$, logarithmic corrections of the form $\log(k L)$ or $\log(a(t))\sim H t$ are allowed by symmetries. 

{\bf $Log(k L)$ effects.}  The factor of $\log(k L)$ can be potentially very large, as $\log(k L)$ is of order $N_{\rm beginning}$, the number of $e$-foldings of Inflation that have occurred {\it before} the mode $k$ has crossed the horizon. Even for the standard inflation that we might have in our past, $N_{\rm beginning}$ can be a large enhancement factor. Furthermore in situations where $N_{\rm beginning}$ might be large, $\langle\zeta^2\rangle$ for modes exiting the horizon at the beginning of inflation might also be significantly larger as one could be near an eternal inflation regime. The infrared factor $\log(k L)$ does appear in the one-loop correction to the power spectrum~\cite{Giddings:2010nc,Byrnes:2010yc}, and in~\cite{projection} some of us have shown that it is simply a projection effect that is completely removed when one computes observable quantities and that does not affect our ability to extract predictions from inflation.

{\bf $Log(a(t))$ effects and the predictivity of Inflation.} In this paper we try to address the question of wether the one-loop correction to the power spectrum is time dependent, or in other words if at loop level $\zeta_k$ is constant after the mode $k$ has crossed the horizon. We notice that  for our current inflationary patch, since we observe around $50$ $e$-foldings of inflation and $\zeta \sim 3\times 10^{-5}$, such a correction factor, even if present, would represent a correction at most of order $50 \times 10^{-9}\sim 5 \times 10^{-8}$. From an observational perspective this is a very small correction. Regardless of this fact, as a matter of principle if such a time-dependent factor were to be present the consequences for the inflationary theory would be profound.  Such a result would imply that short scale fluctuations, say of the size of the horizon, can change the amplitude of a mode after it has crossed the horizon. In standard inflation the amplitude of the short perturbations is very small and the duration of inflation is relatively short so the resulting evolution of the long modes is negligible. However, fluctuations might not be small during other epochs of the evolution of the universe, such as reheating and baryogenesis or if the dynamics of inflation changes dramatically at some point. We know little about these epochs, but if perturbations were to  be large on Hubble scales during those times, the time-dependence induced on long, observable, modes could change their amplitude significantly. We would lose the predictions of Inflation unless we know the details of the physics governing reheating or baryogenesis, which we hardly do. 
%Indeed one of the greatest advantages of Inflation is that its predictions for the density perturbations are independent of the largely unknown history of the universe from the end of Inflation to recombination. This reasoning could completely change if at one loop we were to find that $\zeta$ is time-dependent. 

The potential for a time dependence of the power spectrum at loop level was pointed out by Weinberg in~\cite{Weinberg:2005vy}. He noticed that many diagrams naively induce a time-dependence of $\zeta$~\footnote{Such a result would not be in contradiction with the many proofs available in the literature on the conservation of $\zeta$ outside of the horizon (see for example~\cite{Maldacena:2002vr,Lyth:2004gb,Langlois:2005qp}). The fact that the constant solution is the attractor one, and not simply one of the two solutions, was proven in~\cite{Cheung:2007sv}. All these proofs work in the limit in which all modes are longer than the horizon, so that gradients of all fluctuations can be neglected.}. The question of weather a time dependence persists after we sum all the diagrams has remained open. In \cite{Senatore:2009cf} we addressed this issue in certain simplified examples involving spectator fields running in the loops. Although the physics we identified in that paper will basically apply unchanged in this study, the fact of the matter is that no proper calculation in the context of single clock inflation has been presented.  Ref.~\cite{Kahya:2010xh} claims to have done this and to have found a time dependence. In reality they only presented results for a severely truncated and simplified Lagrangian and of course they did not recover the cancellations we identify in this paper and thus claimed a spurious time dependence. 

{\bf Slow Roll Eternal Inflation.} From a more theoretical point of view, a time-dependence of $\zeta$ would have important consequences for slow-roll eternal inflation. In recent years~\cite{Creminelli:2008es,Dubovsky:2008rf,Dubovsky:2011uy,perko}, there has been remarkable progress in understanding slow roll eternal inflation at a quantitative level. The study of eternal inflation (usually of the false vacuum type) has been largely motivated by the fact that the universe is currently accelerating and by the apparent existence of a landscape of vacua in  String Theory which put together suggest that the current acceleration can be understood as resulting from an anthropic selection of the vacuum energy made possible by an epoch of eternal inflation in our past. Another piece of motivation to study eternal inflation relies on the perhaps mysterious connections between gravity and quantum mechanics in the presence of a horizon. De Sitter space, with its supposedly finite entropy, represents a mystery, and slow roll (eternal) inflation represents a natural regularization of de Sitter space. In~\cite{Creminelli:2008es} it was shown that slow roll inflation undergoes a phase transition when a parameter 
\be
\Omega=\frac{2\pi^2}{3}\frac{\dot\phi^2}{H^4}\ ,
\ee
becomes less than one. At that point, the probability to develop an infinite volume goes from being strictly zero to non-zero. This is the phase transition to eternal inflation. Subsequently, in~\cite{Dubovsky:2008rf}, it was found that there is a sharp upper bound to how large a finite volume can be created: the probability to produce a finite volume larger than $e^{6N_c}$, with $N_c$ representing the classical number of $e$-foldings, is non-perturbatively small from the point of view of quantum gravity:
\be
P\left(V_{\rm finite}> e^{6N_c}\right)< e^{-\mpl^2/H^2}\ .
\ee
By connecting the classical number of $e$-foldings to the the entropy of de Sitter space $S_{dS}$ at the end of inflation, this bound can be recast as 
\be\label{eq:bound}
P\left(V_{\rm finite}> e^{S_{dS}/2}\right)< e^{-\mpl^2/H^2}\ .
\ee
This bound is a generalization to the quantum and eternal regime of the bound found in~\cite{ArkaniHamed:2007ky}, that was much stronger than the one in (\ref{eq:bound}). Further, in~\cite{Dubovsky:2011uy}, it was shown that this bound is actually universal: it holds for any number of spacetime dimensions and for any number of inflating fields. Moreover it holds unchanged also when considering higher-order corrections to the theory of gravity and of the inflaton, and it does so to all orders in perturbation theory. In an upcoming paper~\cite{perko}, some of us will show that it holds also when including slow-roll corrections. All of these results strongly suggest that the bound in (\ref{eq:bound}) is a true fact of nature connected to the holographic interpretation of de Sitter space.

All these new results on Eternal Inflation assumed that the $\zeta$ two-point function at coincidence takes the form\footnote{Studies of the phase transition to slow-roll eternal inflation have only been done at lowest order in slow-roll, where there is basically no distinction between $\zeta$ and $\delta\phi$.}
\be
\langle\zeta(x)^2\rangle\sim H^3 t\ ,
\ee
which is a direct consequence of its scale invariance and time-independence in Fourier space
\be
\langle\zeta_k^2\rangle\sim \frac{H^2}{k^3}\ .
\ee
If the two point function of the inflaton in Fourier space were to go as
\be
\langle\zeta_k^2\rangle\sim \frac{H^2}{k^3}\log(k L) \ , \qquad {\rm or} \qquad \langle\zeta_k^2\rangle\sim \frac{H^2}{k^3} H t
\ee
then in real space it would go as
\be
\langle\zeta(x)^2\rangle\sim H^4 t^2\ ,
\ee
and all the above-mentioned new results on slow roll eternal inflation would fail~\footnote{We acknowledge David Gross for pointing this out to us.}. Depending on the sign of the loop correction, we would be lead to conclude that all inflationary models are either eternal or never-eternal. This motivates us to study the possible time-dependence of $\zeta$ at loop level.

\subsection{Simple Arguments}

There are several simple intuitive arguments that suggest that short scale fluctuations cannot induce a time dependence on a long wavelength $\zeta$ mode that is much longer than the horizon. The simplest and most intuitive argument relies on the fact that at long wavelengths a $\zeta$ mode is indistinguishable in practice from a rescaling of the scale factor $a\rightarrow a\, e^{\zeta}$. This means that a time dependent $\zeta$ is more or less equivalent to a change in the local value of the expansion rate $H$: $\dot\zeta\sim \delta H$. In order for short-scale fluctuations to create a time-dependent long wavelength $\zeta$, the short scale fluctuations should create a modulation of the Hubble parameter that is coherent over a very large scale, the scale of the long wavelength $\zeta$ mode.

One could imagine two mechanisms through which this could happen. The random small scale fluctuations could lead by chance to a large scale fluctuation, but simple `square root of $N$' type of arguments show that this is not the case. Another option is that the short modes are sensitive to the long wavelength fluctuations through tidal-type effects and thus their expectation values, their energy density say, varies over the long scales and leads to a modulation in the expansion rate. This last possibility also sounds quite unreasonable. Because of the attractor nature of the inflationary background, a long wavelength $\zeta$ fluctuation is locally almost indistinguishable from a rescaling of the background, with corrections that rapidly redshift to zero. This means that short wavelength fluctuations should behave in very much the same way in the presence of a long $\zeta$ mode as they do in its absence (apart for a trivial rescaling of the coordinates). This is what the so-called Maldacena consistency condition of curvature fluctuations actually states~\cite{Maldacena:2002vr,Creminelli:2004yq,Cheung:2007sv}, and it has been shown to work at tree-level in several calculations. 

Perhaps a better way to illustrate the point we are trying to make is the following. Assume that short wavelength modes running in the loop lead to a time dependence of the two point function of a long wavelength mode. This one loop calculation is just giving the change of the long modes produced by the short modes when averaged over the short ones. If the short modes can be observed directly the effect of the short modes on the long should lead to an observable correlation between short and long modes. In other words, it should lead for example to a non-zero three point function in the squeezed limit. However, since the work of Maldacena~\cite{Maldacena:2002vr} we know that there is no such effect in the squeezed three point function. It is hard to imagine that one would not be able to detect a correlation between short and long modes when both short and long modes are measured, but that on average the short modes do lead to an evolution of the long modes. 

All of this suggests that it would be quite surprising if short modes were to induce time-dependence in a long wavelength $\zeta$ fluctuation~\footnote{There is one subtlety which has to do with the renormalization of the background. Short wavelength fluctuations do renormalize the background, so that $H(t)$ is different from its value at tree level when the short fluctuations are neglected. It is important to take this fact into account properly in order for $\zeta$ not to have a time-dependence.}.
We note that the essence of these arguments were already given by some of us in~\cite{Senatore:2009cf}.

\subsection{Summary of the Strategy}

Let us make the simple arguments above a bit more precise highlighting our strategy to prove the time-independence of $\zeta$. Since we are interested in a late time-dependence of $\zeta$, we can restrict ourselves to the case in which  we let only short wavelength modes run in the loops. The constancy of $\zeta$ when all modes are outside the horizon was already proved in~\cite{Maldacena:2002vr}. In the present  case, computing one-loop effects can be thought as solving the non-linear evolution equations for a long wavelength $\zeta$ operator, $\zeta_L$, up to cubic order in the fluctuations.  This will take the form
\be
\hat O\left[\zeta_L\right]= S\left[\zeta_S,\zeta_S,\zeta_L\right]\ ,
\ee
where $S$ represent a generic sum of operators that are quadratic in the short wavelength $\zeta$, $\zeta_S$, and that can also eventually depend on $\zeta_L$ both explicitly and implicitly through a dependence of $\zeta_S$ on $\zeta_L$. Each monomial in $S$ can contain derivatives acting on the various $\zeta$'s. The solution is schematically of the form
\be
\zeta_L=\hat O^{-1}\left[S\left[\zeta_S,\zeta_S,\zeta_L\right]\right]\ .
\ee
It should be noted the $\langle S\left[\zeta_S,\zeta_S,\zeta_L\right]\rangle$ is in general not zero. There are tadpole contributions for $\zeta$ because at loop level we are expanding around the incorrect background history. We will add tadpole counterterms to the action to ensure that the background solution we started with satisfies the equations of motion. These counterterms lead to additional diagrams that will cancel many of the one loop diagrams in our power spectrum calculation. 
 
The one loop power spectrum will be given by
\be
\langle\zeta_L\zeta_L\rangle\sim\langle\hat O^{-1}\left[S\left[\zeta_S,\zeta_S,\zeta_L\right]\right]\zeta_L\rangle+\langle\hat O^{-1}\left[S\left[\zeta_S,\zeta_S,\zeta_L=0\right]\right]\hat O^{-1}\left[S\left[\zeta_S,\zeta_S,\zeta_L=0\right]\right]\rangle\ .
\ee
We call the first contribution on the right the cut-in-the-side ($CIS$) diagrams, while the second contribution on the right cut-in-the-middle ($CIM$) diagrams. 

The $CIM$ diagrams represent the effect of the short scale modes in their unperturbed state directly on the power spectrum of the long wavelength modes. These diagrams will not lead to any time-dependence of the long modes simply because it is very hard for short mode fluctuations to be coherent over long scales. 

Many of the $CIS$ diagrams cancel with diagrams coming from the tadpole counterterms. The remaining $CIS$ diagrams represent instead the evolution of $\zeta_L$ due to the effect  that $\zeta_L$ itself has on the expectation value of  quadratic operators made of short modes. These diagrams involve the correlation between this short-mode expectation value and the long wavelength mode itself~\footnote{It will become clear later that this remaining $CIS$ diagrams depend both on the cubic and quartic Hamiltonians.}. This short-mode long-mode correlation sources $\zeta_L$.

The Maldacena consistency condition implies that this short-mode long-mode correlation actually vanishes, 
\be
\langle\hat O^{-1}\left[S\left[\zeta_S,\zeta_S,\zeta_L\right]\right]\zeta_L\rangle=0.
\ee
This is so because the consistency condition means that in the limit in which the long mode has a wavelength much longer than the horizon, it simply acts as a rescaling of the coordinates. So the correlation function between short and long modes can be understood in terms of the power spectrum of the short modes computed in a rescaled background.  Since in the loop the short-mode expectation value is integrated over all the short-mode momenta the rescaling is irrelevant and as a result there is no correlation between the short scale power and the long mode. 

%Once this is done, the remaining operator  $\hat O^{-1}\left[S\left[\zeta_S,\zeta_S,\zeta_L=0\right]\right]$ is precisely the operator that must be cancelled by imposing that the tadpoles for $\zeta$ cancel. This shows that there are no sources left for a long wavelength $\zeta_L$ when it is very outside of the horizon. It proves that its correlation function becomes constant in time.

Even though the former arguments are quite compelling, the calculation is very complex, and many subtleties are hidden in the above equations. They include the identification of the Lagrangian of the $\zeta$ zero-mode, that will turn out to be delicate and to affect the definition of the tadpole counterterms.  Because of diff. invariance, these counterterms will play a role even for the finite momentum correlation functions. Additionally, it will be non-trivial to see how the Maldacena consistency condition  works when dealing with operators involving derivatives. 

In summary, since the interactions are dominated by the gravitational ones, our one-loop computation amounts to doing a one loop calculation in gravity in an accelerating universe. This is quite a hard task, at least for us!  In particular, there are {\it many many many} diagrams involved, and {\it many many} of these naively induce a time-dependence on $\zeta$. The time independence will result from cancellations among diagrams. We will now try to move step by step to make our arguments explicit and precise, finally proving that $\zeta$ is constant outside of the horizon also at one-loop level.

%{\bf ??!!?!?!?!?!?!?!!? Shall we put discussion on constrained variables?  ?!!?!?!?!?!?!?!?!?!!?!?!?!?!?!?}
%
%{\bf ????!?!!?!?!?!? Further I did not mentioned here or in the conclusions the fact that the zero mode and the finite k mode are treated in a slightly different way. I guess it is ok, but let me know if you want to put it in here or in the conclusions as well???????/}

\section{An Intuitive Organization of the Diagrams}

It is possible to organize the one-loop diagrams in a way that is particularly close to our intuition. 
   This approach was originally developed in \cite{Musso:2006pt} for a restricted set of theories, and it was noted in~\cite{Adshead:2008gk} that the derivation was not consistent with the $i\,\epsilon$ prescription for choosing the interacting vacuum in the past. This approach has been generalized in~\cite{Senatore:2009cf} to more generic theories and a  correct $i\,\epsilon$ prescription has been implemented. Here we will see that the implementation of the $i\,\epsilon$ prescription can be performed in a very simple way.

For concreteness let us specialize to the $\zeta$ two-point function. We have to compute
\be\label{eq:generic_expectation}
\langle\Omega|\zeta^2(t)|\Omega\rangle=\langle0| U_{int}(t,-\infty_+)^{\dag} \zeta^2_I(t) U_{int}(t,-\infty_+) |0\rangle\ ,
\ee
where $|\Omega\rangle$ is the vacuum of the interacting theory, $|0\rangle$ is the one of the free theory,
\be
U_{int}(t,-\infty_+)=T e^{-i \int_{-\infty_+}^tdt'\;  H_{int}(t')}\ ,
\ee
and the subscript $_I$ stays for interaction picture. Finally, the symbol $-\infty_+$ represents the fact that the time-integration contour has been rotated so as to project the free vacuum on the interacting vacuum. In practice, this amounts to choosing the contour that suppresses the oscillatory terms in the infinite past.

We start by taking expression (\ref{eq:generic_expectation}) and inserting the unit operator
\be
1=U_{int}(t,-\infty) U^\dag_{int}(t,-\infty)\ , 
\ee
between the two $\zeta$'s, to obtain
\be\label{eq:generic_expectation_new}
\langle\zeta^2(t)\rangle=\langle\left( U^\dag_{int}(t,-\infty_-) \zeta_I(t) U_{int}(t, -\infty) \right)\left(U^\dag_{int}(t,-\infty) \zeta_I(t)U_{int}(t,-\infty_+) \right)\rangle\ ,
\ee
where we have ignored to specify the state upon which we compute the correlation function, either $|\Omega\rangle$ or $|0\rangle$, as it is clear from the context. Ignoring for a moment the issue of the $i\, \epsilon$ prescription, we have written the expectation of the operator $\zeta(t)^2$ as the product of two interaction picture $\zeta_I(t)$'s, each evolved with the interaction picture time evolution operator $U_{int}$. In other words, the $\zeta(t)^2$ correlation function is simply given by the correlation function of the evolved $\zeta(t)$'s. 
%A closer look at (\ref{eq:generic_expectation_new}) might let us think that this picture does not quite work because of the difference in the extremes and the contour of integration of the time integrals in the tarious evolutors. However, the  
We can Taylor expand in $H_{int}$ to obtain
\bea\label{eq:generic_expectation_new_expanded}
&&\langle\zeta^2(t)\rangle=\\ \nonumber
&&=\langle\left(\sum_{N=0}^\infty i^N\int^t_{-\infty} dt_N\int_{-\infty} ^{t_N}dt_{N-1}\ldots \int_{-\infty} ^{t_2}dt_1 \left[H_{int}(t_1),\left[H_{int}(t_2),\ldots  \left[H_{int}(t_N),\zeta_I(t)\right]\ldots\right]\right]\right)\\ \nonumber 
&&\times \left(\sum_{N=0}^\infty i^N\int_{-\infty} ^t dt'_N\int_{-\infty} ^{t_N}dt'_{N-1}\ldots \int_{-\infty} ^{t_2}dt'_1 \left[H_{int}(t'_1),\left[H_{int}(t'_2),\ldots  \left[H_{int}(t'_N),\zeta_I(t)\right]\ldots\right]\right]\right)^\dag\rangle\  .
\eea
Expanding (\ref{eq:generic_expectation_new_expanded}) up to second order in $H_{int}$, we obtain
\be
\langle\zeta^2(t)\rangle=\langle\zeta^2(t)\rangle_{CIS}+\langle\zeta^2(t)\rangle_{CIM}\ ,
\ee
where we  have defined
\bea\nonumber\label{eq:CIMandCIS}\nonumber
\langle\zeta^2(t)\rangle_{CIS}&=&-2\,{\rm Re}\left[\left( \int^{t}_{-\infty} dt_{2} \int^{t_2}_{-\infty}dt_1 \langle\left[H_{3}(t_1),\left[H_{3}(t_2),\zeta_I(t)\right]\right]\right)\zeta_I(t)\rangle\right.\\  \label{eq:CIS_diagram}
&&\left.\qquad \qquad - i \left( \int^{t}_{-\infty}dt_{1} \langle\left[H_{4}(t_1),\zeta_I(t)\right]\right)\zeta_I(t)\rangle\right]\ , \nonumber \\  \label{eq:CIM_diagram}
\langle\zeta^2(t)\rangle_{CIM}&=&-\left( \int^{t}_{-\infty}dt_1\langle \left[H_{3}(t_1),\zeta_I(t)\right]\right)\left( \int^{t}_{-\infty}dt'_1 \left[H_{3}(t'_1),\zeta_I(t)\right]\rangle\right)\ .
\eea
The subscript $_{CIS}$ denotes what we call cut-in-the-side diagrams, while $_{CIM}$ denotes cut-in-the-middle diagrams.  Here by $H_3,\,H_4,\ldots$ we mean the cubic, quartic, $\ldots$ interaction Hamiltonians.
We see that the $CIM$ diagrams are made up by evolving each of the two $\zeta$'s to first order in the cubic interactions. The $CIS$ diagrams corresponds to evolving only one of the two $\zeta$'s, either twice with cubic interactions or once with a quartic interaction. 
%This explains the name we give to this diagrams.

\begin{figure}
\begin{center}
\includegraphics[width=10cm]{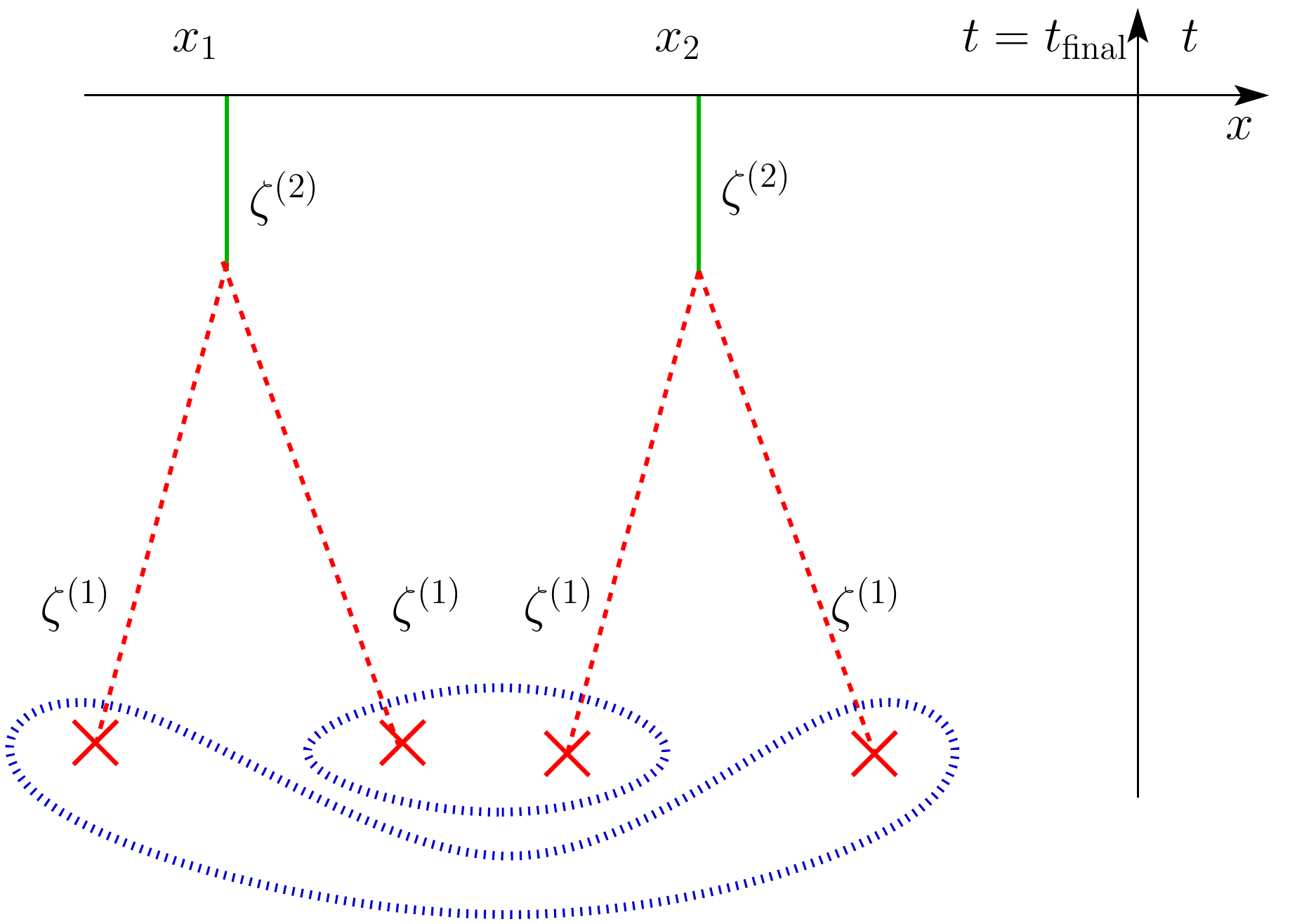}
\caption{\label{CIM} \small\it Cut-in-the-middle ($CIM$) diagrams. Green continuos lines represent Green's functions, red dashed lines represent free fields, and red crosses circled by a blue dotted line represent correlations of free fields. Two crosses have to be contracted together in order for the diagram not to be zero. This diagram represents how vacuum correlation functions of quadratic operators $\zeta^{(1)}{}^2$, $\langle\zeta^{(1)}{}^2\zeta^{(1)}{}^2\rangle$ source perturbed correlation functions for $\zeta^{(2)}$: $\langle\zeta^{(2)}\zeta^{(2)}\rangle$.}
\end{center}
\end{figure}

This form of organizing the diagrams is particularly intuitive.  If we remind ourselves that the $\zeta$ retarded Green's function is given by
\be
G^R_\zeta(x,x')=i\theta(t-t')\left[\zeta_I(x),\zeta_I(x')\right]\ ,
\ee
we have that 
\be
[H_{int}^{(3)},\zeta]\sim G^R \frac{\delta {\cal L}_3}{\delta \zeta}\ .
\ee
Then the $CIM$ diagrams  approximately correspond to considering the sourcing of $\zeta$ from the vacuum correlation function of $\delta {\cal L}_3/\delta \zeta$. This is very similar to the case when we try to solve some equations of motion perturbatively. We can define the solution of order $n$ in the perturbation as $\zeta^{(n)}$. If we have schematically:
\be
{\cal D}\zeta^{(2)}=\zeta^{(1)}{}^2 \qquad\Rightarrow \qquad \zeta^{(2)}=\int dt' G_\zeta(t,t')\zeta^{(1)}(t'){}^2
\ee
where $\cal D$ is the differential operator of the free equations of motion, of which the Green's function is the inverse, then the CIM diagram is represented by the following
\be
CIM=\langle\zeta^{(2)}\zeta^{(2)}\rangle\  .
\ee
The $CIM$ diagram is diagrammatically represented in Fig.~\ref{CIM}. Intuitively, it can be thought of as taking into account of the backreaction on $\zeta$ from the quantum variance of the operator $\delta {\cal L}_3/\delta \zeta$.
 
\begin{figure}
\begin{center}
\includegraphics[width=10cm]{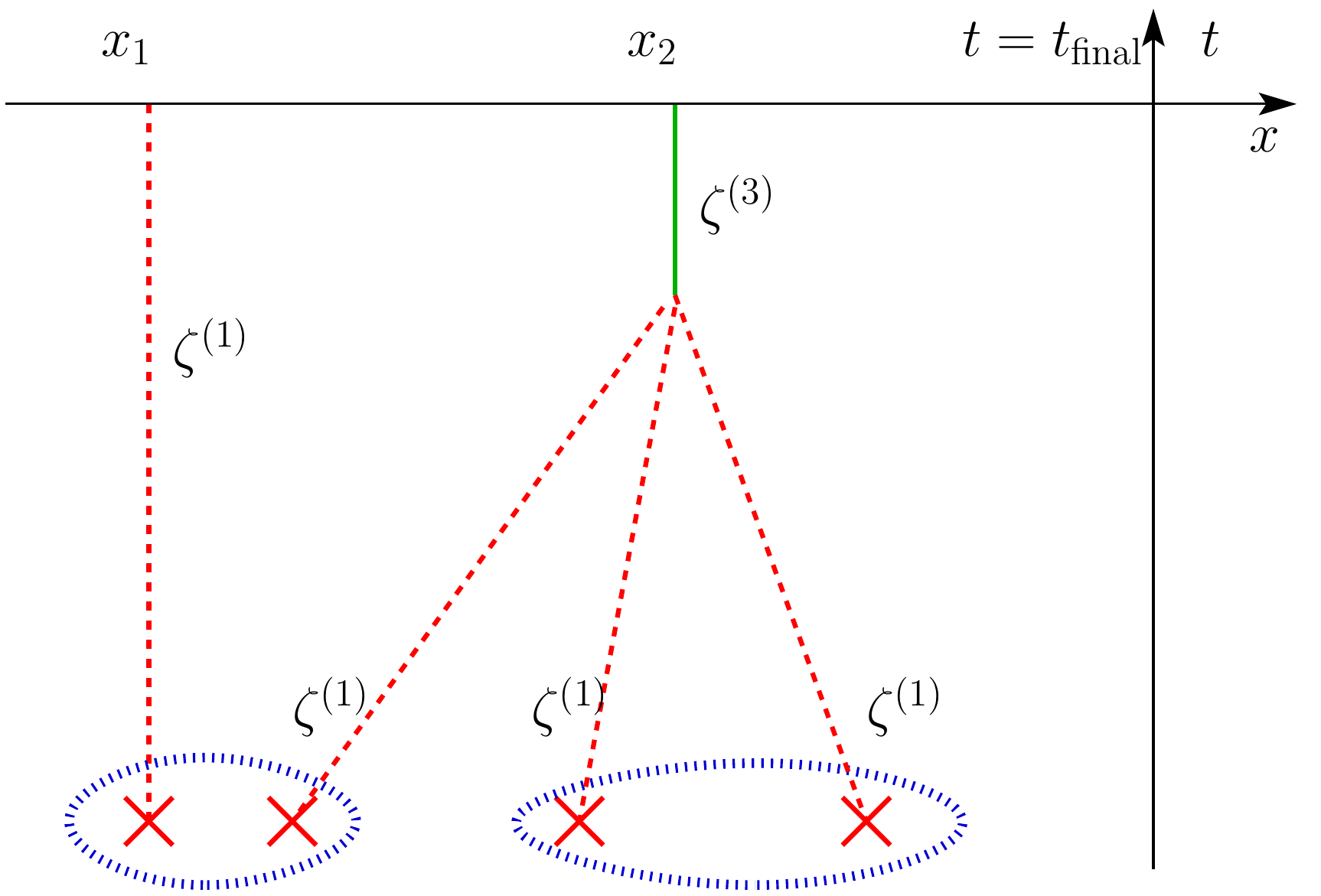}
\caption{\label{CIS_quartic} \small\it Cut-in-the-side quartic ($CIS_4$) diagrams. These diagrams represent how vacuum expectation values of quadratic operators $\langle\zeta^{(1)}{}^2\rangle$ affect the propagation of a mode $\zeta^{(3)}$, and therefore the $\zeta$ correlation function: $\langle\zeta^{(3)}\zeta^{(1)}\rangle$}
\end{center}
\end{figure}

On the other hand, the $CIS$ diagrams correspond to two sort of diagrams. The ones involving the quartic interactions, $CIS_4$, correspond to considering the effect of the expectation value of the vacuum fluctuations of two fluctuations on the external $\zeta$. Schematically, it is given by
\be
{\cal D}\zeta=\zeta^{(1)}{}^3\qquad\Rightarrow \qquad \zeta^{(3)}=\int dt' G_\zeta(t,t')\zeta^{(1)}(t'){}^3\qquad \Rightarrow\qquad CIS_4=\langle\zeta^{(3)}\zeta^{(1)}\rangle\  , 
\ee 
and it is represented in Fig.~\ref{CIS_quartic}. %Its physical intuition is quite more complex, and it will be discussed later.

\begin{figure}
\begin{center}
\includegraphics[width=10cm]{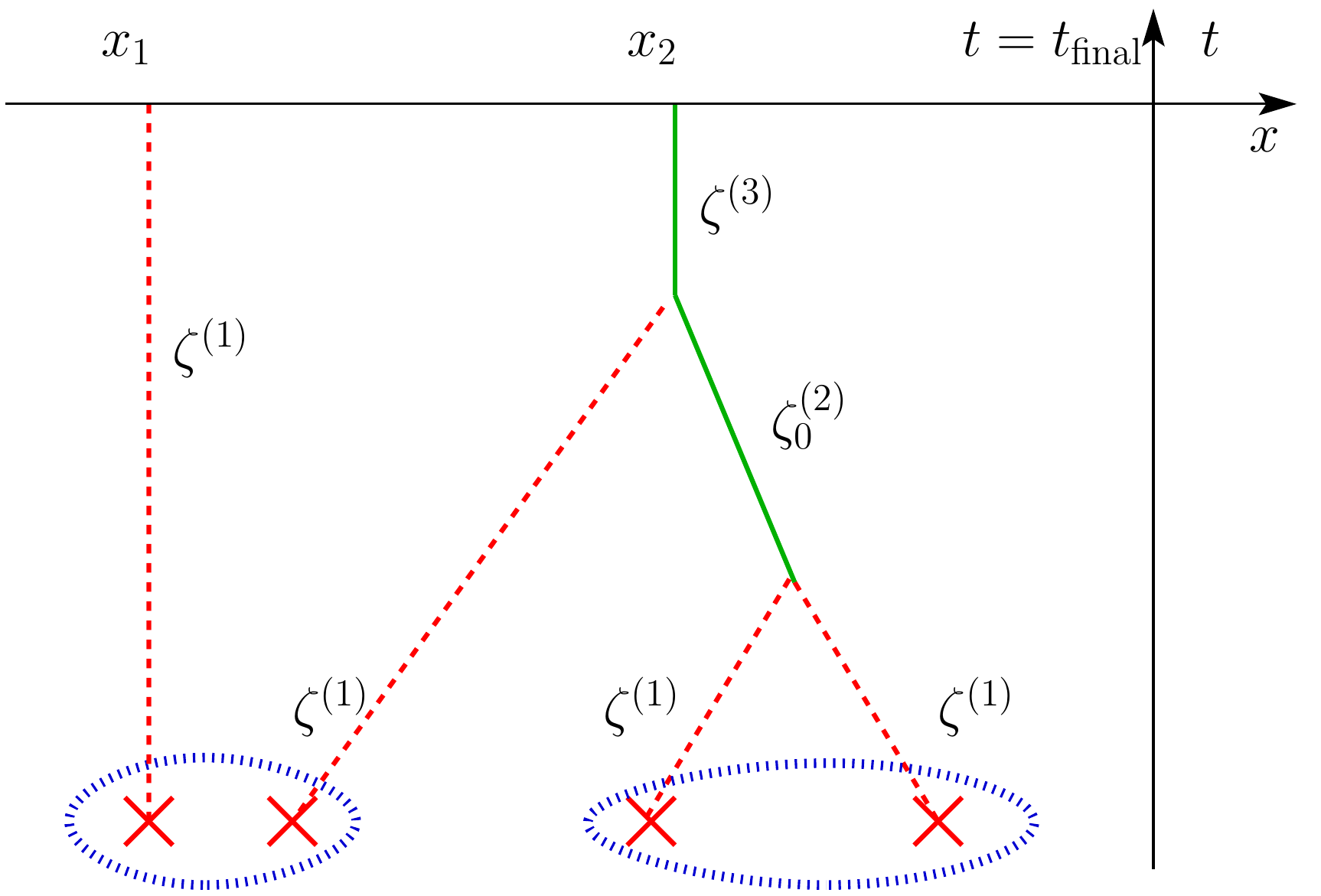}
\caption{\label{CIS_non_1PI} \small\it Non-1PI cut-in-the-side quartic ($CIS_{non-1PI}$) diagrams. These diagrams represent how vacuum expectation values of quadratic operators $\langle\zeta^{(1)}{}^2\rangle$ affect the propagation of the zero mode $\zeta^{(2)}_0$, and therefore the evolution of a mode by a non linear coupling $\zeta^{(3)}\sim \zeta^{(1)}\zeta_0^{(2)}$. This sources a correlation function of the form: $\langle\zeta^{(3)}\zeta^{(1)}\rangle$}
\end{center}
\end{figure}

The $CIS$ diagrams that  involve two cubic interactions can in turn be divided in two subclasses. The first are of the non-1PI form, $CIS_{non-1PI}$, and describe the effect of the expectation value of two fluctuations on the $\zeta$ zero mode, $\zeta_0$, and how then the zero mode affects the $\zeta$ propagation. Schematically, this is given by
\bea
&&\!\!\!\!\!\!\!\!\!\!\!{\cal D}\zeta_0=\zeta^{(1)}{}^2\qquad\Rightarrow \qquad \langle\zeta^{(2)}_0\rangle=\int dt' G_\zeta(t,t')\langle\zeta^{(1)}(t')^2\rangle{}\qquad\\ \nonumber
&&\!\!\!\!\!\!\!\!\!\! {\cal D}\zeta^{(3)}=\zeta^{(1)}\langle\zeta_0^{(2)}\rangle \qquad\Rightarrow \qquad \zeta^{(3)}=\int dt' G_\zeta(t,t')\zeta^{(1)}(t')\langle\zeta_0^{(2)}\rangle(t')\qquad \Rightarrow\qquad  CIS_{non-1PI}=\langle\zeta^{(3)}\zeta^{(1)}\rangle\; ,
\eea
and it is represented in  Fig.~\ref{CIS_non_1PI}.  This diagram intuitively represents how a perturbation to the background (the zero mode) affects the evolution of the finite-$k$ modes. 

\begin{figure}
\begin{center}
\includegraphics[width=10cm]{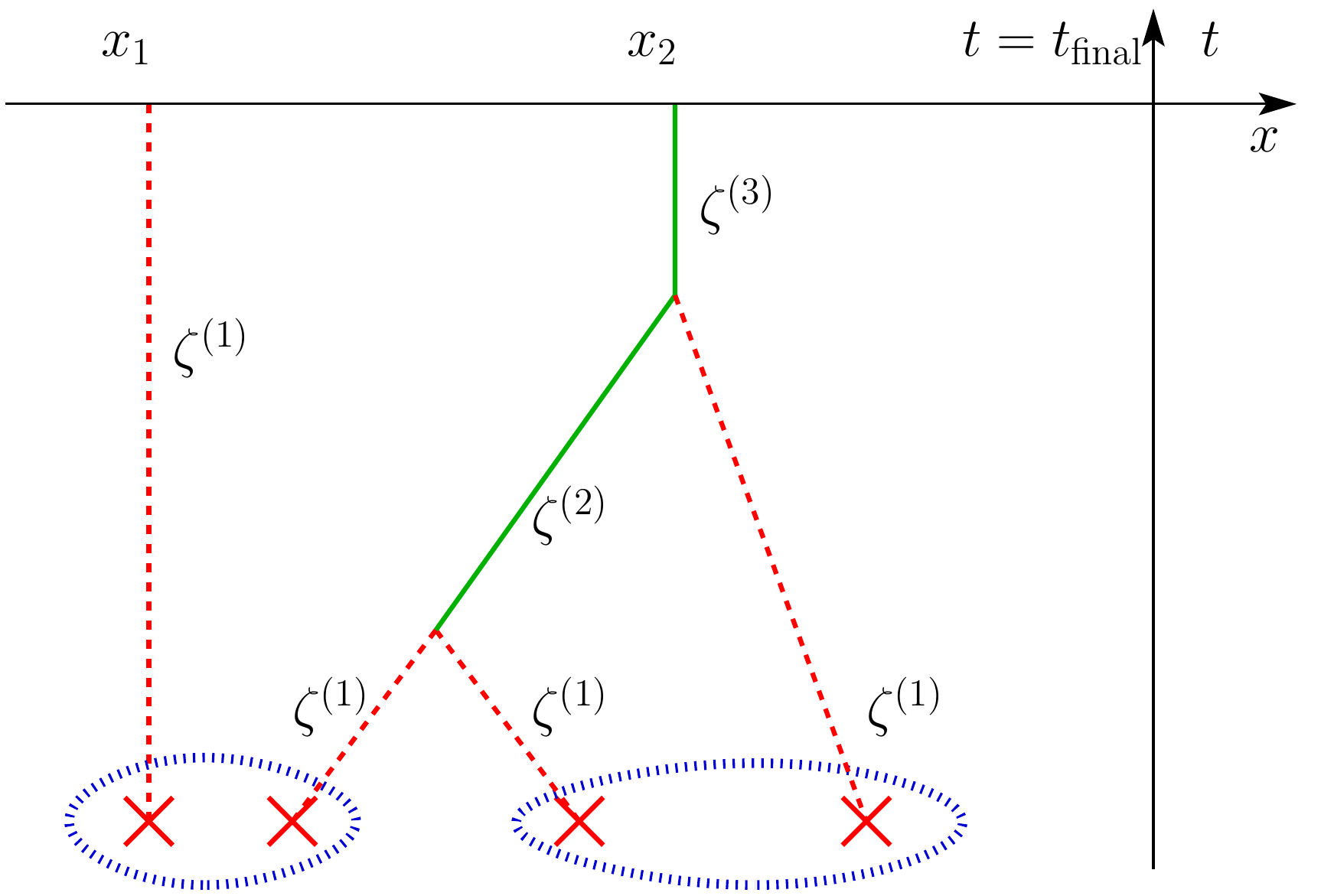}
\caption{\label{CIS_1PI} \small\it 1PI cut-in-the-side quartic ($CIS_{1PI}$) diagrams. These diagrams represent how the propagation of a mode is perturbed at two different times by two fluctuations that are correlated among themselves. This sources a correlation function of the form: $\langle\zeta^{(1)}\zeta^{(3)}\rangle$}
\end{center}
\end{figure}

The second kind of $CIS$ diagram, $CIS_{1PI}$ is 1PI and corresponds to considering the sourcing on $\zeta$ from two fluctuations, one of which has been perturbed by an initial $\zeta$ fluctuation.
\bea
&&\!\!\!\!\!\!\!\!\!\! {\cal D}\zeta^{(2)}=\zeta^{(1)}{}^2\qquad\Rightarrow \qquad \zeta^{(2)}=\int dt' G_\zeta(t,t')\zeta^{(1)}(t'){}^2\qquad\\ \nonumber
&&\!\!\!\!\!\!\!\!\!\! {\cal D}\zeta^{(3)}=\zeta^{(1)}\zeta^{(2)} \qquad\Rightarrow \qquad \zeta^{(3)}=\int dt' G_\zeta(t,t')\zeta^{(1)}(t')\zeta^{(2)}(t')\qquad \Rightarrow\qquad  CIS_{1PI}=\langle\zeta^{(3)}\zeta^{(1)}\rangle\; ,
\eea
and it is represented in Fig.~\ref{CIS_1PI}. Physically, this represents how a fluctuation is affected by two fluctuations, one of which has been already perturbed. If we imagine for a moment that only short fluctuations run in the loop, this diagram would represent how a long mode affects through tidal effects the dynamics of the short modes, and how these backreact on the long mode.

Let us finally comment on how to implement the $i\,\epsilon$ prescription. When we insert the unit operator in (\ref{eq:generic_expectation_new}), we should keep in mind that the integration contours of the time evolutors on the sides of the expectation value are rotated, while the ones in the middle are not. This means that when we Taylor expand in $H_{int}$, the various terms do not really regroup and form commutators,  because they are evaluated on different contours. A solution to this problem was provided in~\cite{Senatore:2009cf} where the rotation was performed only at very early times and the commutator form applied only at late time. Here we implement the correct $i\,\epsilon$ rotation in a different way. We perform no contour rotation, but we multiply our expression by $e^{i\epsilon (\sum k_i) t}$, where the sum runs over all the momenta involved in the loops and $\epsilon>0$, so that the time integrals are convergent in the far past, and then take the limit $\epsilon\rightarrow 0^+$. While the multiplication by $e^{i\epsilon (\sum k_i) t}$ is not a rotation of the contour of integration, it converges to one in the limit $\epsilon\rightarrow 0^+$. It can be easily checked that this procedure agrees with the rotation of the contour.

%\begin{figure}
%\begin{center}
%\includegraphics[width=7cm]{CIM_zeta.pdf}
%\caption{\label{CIM} \small\it Cut-in-the-middle ($CIM$) diagrams. Continuos lines represents Green's functions, dashed lines represent free fields, and crosses represent correlations of free fields. Two crosses have to be contracted together in order for the diagram not to be zero. This diagram represents how vacuum correlation function of quadratic operators $\zeta^{(1)}{}^2$, $\langle\zeta^{(1)}{}^2\zeta^{(1)}{}^2\rangle$ source perturbed correlation functions for $\zeta^{(2)}$: $\langle\zeta^{(2)}\zeta^{(2)}\rangle$.}
%\end{center}
%\end{figure}

%\begin{figure}
%\begin{center}
%\includegraphics[width=10cm]{CIS_zeta_quartic.pdf}
%\caption{\label{CIS_quartic} \small\it Cut-in-the-side quartic ($CIS_4$) diagrams. These diagrams represent how vacuum expectation values of quadratic operators $\langle\zeta^{(1)}{}^2\rangle$ affect the propagation of a mode $\zeta^{(3)}$, and therefore the $\zeta$ correlation function: $\langle\zeta^{(3)}\zeta^{(1)}\rangle$}
%\end{center}
%\end{figure}

%\begin{figure}
%\begin{center}
%\includegraphics[width=10cm]{CIS_zeta_non_1PI.pdf}
%\caption{\label{CIS_non_1PI} \small\it Non-1PI cut-in-the-side quartic ($CIS_{non-1PI}$) diagrams. These diagrams represent how vacuum expectation values of quadratic operators $\langle\zeta^{(1)}{}^2\rangle$ affect the propagation of the zero mode $\zeta^{(2)}_0$, and therefore the evolution of a mode by a non linear coupling $\zeta^{(3)}\sim \zeta^{(1)}\zeta_0^{(2)}$. This sources a correlation function of the form: $\langle\zeta^{(3)}\zeta^{(1)}\rangle$}
%\end{center}
%\end{figure}

%\begin{figure}
%\begin{center}
%\includegraphics[width=11cm]{CIS_zeta_1PI.pdf}
%\caption{\label{CIS_1PI} \small\it 1PI cut-in-the-side quartic ($CIS_{non-1PI}$) diagrams. These diagrams represent how the propagation of a mode is perturbed at two different times by two fluctuations that are correlated among themselves. This sources a correlation function of the form: $\langle\zeta^{(3)}\zeta^{(1)}\rangle$}
%\end{center}
%\end{figure}

\section{Loops as the integral of the three-point function}

Let us consider a cubic Lagrangian of the form 
\be
{\cal L}_3=\sum_n {\cal L}_3^{(n)}
\ee
where the sum over $n$ runs over all possible monomials constituting ${\cal L}_3$. We will schematically write
\be
 {\cal L}_3^{(n)} \propto  {\cal D}^{(n)}_1 \zeta {\cal D}^{(n)}_2 \zeta {\cal D}^{(n)}_3 \zeta,
\ee
where ${\cal D}^{(n)}_a$, $a=1,2,3$ are the differential operator acting on $\zeta(x)$ in position $a$. It includes both time and spatial derivatives, as well as the identity operator.

There are certain quartic diagrams which we call $Quartic_{3,\d_t}$. They are the quartic diagrams with the quartic vertices that arise because the cubic Lagrangian contains $\dot\zeta$, $H_4\supset H_{4,3^2}= \delta \dot\zeta/\delta P\times (\delta{\cal L }_3/\delta \dot\zeta)^2/2$.
We want to prove that we can write the sum of $CIS_{1PI}+CIM+Quartic_{3,\d_t}$ diagrams as:
\bea\label{eq:master}
&&\langle\zeta_k\zeta_{k}\rangle_{CIS_{1PI}+CIM+Quartic_{3,\d_t}}=\lim_{\epsilon\rightarrow 0}\int_{-\infty}^t dt_1\; a(t_1)^{3+\delta}\\ \nonumber 
&&\qquad\sum_{a,n}{\cal D}_a^{(n)}G_{\zeta_k}(t,t_1) 2{\rm Re}\langle\left[\frac{1}{a(t_1)^{3+\delta}}\frac{\delta{\cal L}_3^{(n)}(t_1)}{\delta {\cal D}^{(n)}_a\zeta_a(t_1)}\right]_kU_{int}^\dag(t_1,-\infty)\zeta_{k,I}(t)U_{int}(t_1,-\infty)\rangle \; e^{\epsilon k t_1}\ .
\eea
In this formula
 \be
\left[\frac{1}{a(t_1)^{3+\delta}}\frac{\delta{\cal L}_3^{(n)}(t_1)}{\delta {\cal D}^{(n)}_a\zeta_a(t_1)}\right]_k
\ee
represents the $k$-Fourier component of what is left of the  the cubic Lagrangian term ${\cal L}^{(n)}_3$ after the removal of $a(t_1)^{3+\delta}\,{\cal D}^{(n)}_a\zeta_a(t_1)$. $\zeta_I$ is again the interaction picture field.

%The $i\epsilon$ rotation is most simply defined as adding an exponential damping term of the form ${\rm exp}(\epsilon(q_1+q_2+k)t)$ in every time integral, and then sending  $\epsilon\rightarrow 0$. This has been Mathematica(lly) proved.

Eq.~(\ref{eq:master}) is a remarkably simple formula given that it sums up a very large number of diagrams. It shows that the sum of all these diagrams can be written as a sum of integrals of three-point functions. Since we are interested in the case in which the fluctuations running in the loop are much shorter-wavelength than the one in the external fields, the three-point functions are computed in the squeezed limit, a fact that simplifies largely their behavior and makes them describable using the consistency condition of three-point functions. This will turn out to be very useful.

\subsection{Quasi 3-point function}
In order to prove the master eq.~(\ref{eq:master}), let us start by considering the 3-point function appearing there:
\bea\nonumber
&&2{\rm Re}\langle\left[\frac{1}{a(t_1)^{3+\delta}}\frac{\delta{\cal L}^{(n)}_3(t_1)}{\delta {\cal D}_a^{(n)}\zeta_a(t_1)}\right]_kU_{int}^\dag(t_1,-\infty)\zeta_{I,k}(t)U_{int}(t_1,-\infty)\rangle=\\ \nonumber
&&2{\rm Re}\left\{\langle U_{int}(t_1,-\infty)^\dag\left[\frac{1}{a(t_1)^{3+\delta}}\frac{\delta {\cal L}_3^{(n)}(t_1)}{\delta {\cal D}^{(n)}_a \zeta_a(t_1)}\right]_{I,k}U_{int}(t_1,-\infty)U_{int}^\dag(t_1,-\infty)\zeta_{I,k}(t)U_{int}(t_1,-\infty)\rangle+\right.\\ \nonumber
&&\qquad\left.\sum_b\frac{i}{2}\langle\left[{\cal D}^{(n,out)}_b\left[H_3(t_1),\zeta(t_1)\right]\left(\frac{1}{a(t_1)^{3+\delta}}\frac{\delta^2{\cal L}^{(n)}_3(t_1)}{\delta {\cal D}^{(n)}_a\zeta_a(t_1)\delta \dot\zeta_b(t_1)}\right)+\right.\right.\\
&&\qquad\quad\ \ \left.\left.\left(\frac{1}{a(t_1)^{3+\delta}}\frac{\delta^2{\cal L}_3^{(n)}(t_1)}{\delta {\cal D}_a^{(n)}\zeta_a(t_1)\delta \dot\zeta_b(t_1)}\right){\cal D}^{(n,out)}_b\left[H_3(t_1),\zeta(t_1)\right]\right]_k \zeta_k(t) \rangle\right\}\ ,
\eea
where
\be
\left(\frac{\delta^2{\cal L}^{(n)}_3(t_1)}{\delta {\cal D}_a^{(n)}\zeta_a(t_1)\delta \dot\zeta_b(t_1)}\right)
\ee
represents the removal of $\dot\zeta(t_1)$ in position $b$ from the quadratic term $\left(\delta{\cal L}_3^{(n)}(t_1)/\delta {\cal D}^{(n)}_a\zeta_a(t_1)\right)$. Finally ${\cal D}^{(n,out)}_b{}$ is the derivative operator acting on $\zeta_b$ outstripped of the time derivative. For example if ${\cal D}_b\zeta_b=\d\dot\zeta_b$, then ${\cal D}^{(out)}_b=\d$.
%represents the removal of ${\cal D}_i\zeta(t_1)$ from the quartic term $H_{4,3^2}=\left(\delta{\cal L}_3(t_1)/\delta P(t_1)\right)^2$.
The last contact terms are due to the fact that $\dot\zeta$ is not the momentum conjugate to $\zeta$. The simplest way to obtain its time evolution is  using
$\d_t(U_{int}^\dag(t)\zeta_I(t) U_{int}(t))$. When the time derivative acts on the $U_{int}$s it results in contact terms.  We have also symmetrized its expression because ${\cal L}_3$ is hermitian. 
Straightforward manipulations lead~to
\bea
&&\!\!\!\!\!\!\!\!\!\!\!\!2{\rm Re}\langle\left[\frac{1}{a(t_1)^{3+\delta}}\frac{\delta{\cal L}^{(n)}_3(t_1)}{\delta {\cal D}_a^{(n)}\zeta_a(t_1)}\right]_kU_{int}^\dag(t_1,-\infty)\zeta_{I,k}(t)U_{int}(t_1,-\infty)\rangle=\\ \label{eq:piece1}
&&\!\!\!\!\!\!\!\!\!\!\!\!2{\rm Re}\left\{\langle \left[ i \int^{t_1}_{-\infty}dt_2\;H_3(t_2),\frac{1}{a(t_1)^{3+\delta}}\frac{\delta{\cal L}^{(n)}_3(t_1)}{\delta {\cal D}^{(n)}_a\zeta_a(t_1)}\right]_k\zeta_k(t)\rangle\right. \\ \label{eq:piece2}
&&\!\!\!\!\!\!\!\!\!\!\!\!+\sum_{m,b}\langle\left[\frac{1}{a(t_1)^{3+\delta}}\frac{\delta{\cal L}_3^{(n)}(t_1)}{\delta {\cal D}^{(n)}_a\zeta_a(t_1)}\right]_k\int_{-\infty}^{t_1}dt_2\; {\cal D}^{(m)}_bG_{\zeta_k}(t,t_2)\left[\frac{\delta{\cal L}^{(m)}_3(t_2)}{\delta \tilde{\cal D}^{(m)}_b\zeta_b(t_2)}\right]_k\rangle\\ \label{eq:piece3}
&&\!\!\!\!\!\!\!\!\!\!\!\!\!\!-\frac{1}{2}\left.\sum_{b}\langle\left[{\cal D}^{(n,out)}_b\left(\frac{\delta\tilde{\cal L}_3(t_1)}{\delta P(t_1)}\right)\left(\frac{1}{a(t_1)^{3+\delta}}\frac{\delta^2{\cal L}^{(n)}_3(t_1)}{\delta {\cal D}^{(n)}_a\zeta_a(t_1)\delta \dot\zeta_b(t_1)}\right)\right.\right.\\ \nonumber 
&&\qquad\left.\left.+\left(\frac{1}{a(t_1)^{3+\delta}}\frac{\delta^2{\cal L}^{(n)}_3(t_1)}{\delta {\cal D}_a^{(n)}\zeta_a(t_1)\delta \dot\zeta_b(t_1)}\right){\cal D}^{(n,out)}_b\left(\frac{\delta\tilde {\cal L}_3(t_1)}{\delta P(t_1)}\right)\right]_k \zeta_k(t) \rangle\right\}\nonumber \ ,
\eea
where
\be
G_{\zeta}(t,t_1)=i\theta(t-t_1)[\zeta(t),\zeta(t_1)]
\ee
is the $\zeta$ Green's function from $t_1$ to $t$.
The second term is obtained upon noticing that
\be
[\int^t_{-\infty} dt_2\;H_3(t_2),\zeta_k(t)]=-i \int^t_{-\infty} dt_2\; \sum_{m,b}{\cal D}^{(m)}_bG_{\zeta_k}(t,t_2)\left[\frac{\delta{\cal L}^{(m)}_3(t_2)}{\delta{\cal D}^{(m)}_b\zeta_b(t_2)}\right]_k\ ,
\ee
and the third term through the following
\be
\left[H_3(t_1),\zeta(t_1)\right]=-i \frac{\delta \tilde H_3}{\delta P}=i \frac{\delta \tilde {\cal L}_3}{\delta P}\ ,
\ee
where $P$ is the momentum conjugate to $\zeta$ in the interaction picture: $P=\delta{\cal L}_2/\delta\dot\zeta$,  and  we introduced $\tilde H_3$ because any additional  (spatial) derivatives acting on $P$ have been integrated by parts and now act on $H_3$. 
Let us label the term in line (\ref{eq:piece1}) by ${\cal I}_1^{(n)}$, the one in line (\ref{eq:piece2}) by ${\cal I}_2^{(n)}$ and the one in line (\ref{eq:piece3}) by ${\cal I}_3^{(n)}$:
\bea
{\cal I}_1^{(n,a)}(t_1)&=&2{\rm Re}\left\{\langle\left[i \int^{t_1}_{-\infty}dt_2\;H_3(t_2),\frac{1}{a(t_1)^{3+\delta}}\frac{\delta{\cal L}^{(n)}_3(t_1)}{\delta {\cal D}^{(n)}_a\zeta_a(t_1)}\right]_k\zeta_k(t)\rangle\right\}\ , \\ \nonumber
{\cal I}_2^{(n,a)}(t_1)&=&2\sum_{m,b}{\rm Re}\left\{\langle\left[\frac{1}{a(t_1)^{3+\delta}}\frac{\delta{\cal L}^{(n)}_3(t_1)}{\delta {\cal D}^{(n)}_a\zeta_a(t_1)}\right]_k\int_{-\infty}^{t_1}dt_2\;{\cal D}^{(m)}_bG_{\zeta_k}(t,t_2)\left[\frac{\delta{\cal L}^{(m)}_3(t_2)}{\delta \tilde{\cal D}^{(m)}_b\zeta_b(t_2)}\right]_k\rangle\right\}\ , \\ \nonumber
{\cal I}_3^{(n,a)}(t_1)&=&-\sum_{b}{\rm Re}\left\{\langle\left[{\cal D}^{(n,out)}_b\left(\frac{\delta\tilde {\cal L}_3(t_1)}{\delta P(t_1)}\right)\left(\frac{1}{a(t_1)^{3+\delta}}\frac{\delta^2{\cal L}^{(n)}_3(t_1)}{\delta {\cal D}^{(n)}_a\zeta_a(t_1)\delta \dot\zeta_b(t_1)}\right)\right.\right.\\ \nonumber
&&\qquad\qquad\left. \left.+\left(\frac{1}{a(t_1)^{3+\delta}}\frac{\delta^2{\cal L}^{(n)}_3(t_1)}{\delta {\cal D}^{(n)}_a\zeta_a(t_1)\delta \dot\zeta_b(t_1)}\right){\cal D}^{(n,out)}_b\left(\frac{\delta\tilde{\cal L}_3(t_1)}{\delta P(t_1)}\right)\right]_k \zeta_k(t) \rangle\right\}\ ,
\eea
so that
\bea
2{\rm Re}\langle\left[\frac{1}{a(t_1)^{3+\delta}}\frac{\delta{\cal L}^{(n)}_3(t_1)}{\delta {\cal D}^{(n)}_a\zeta_a(t_1)}\right]_k\zeta_k(t)\rangle={\cal I}_1^{(n,a)}(t_1)+{\cal I}_2^{(n,a)}(t_1)+{\cal I}_3^{(n,a)}(t_1)\ .
\eea
 We are now going to see that the $CIS$ diagrams reduce to the sum over $a$ and $n$ of the integral of the  Green's function times ${\cal I}_1^{(n,a)}$, the $CIM$ diagrams reduce to the integral of  Green's function or of its derivatives times ${\cal I}_2^{(n,a)}$, and finally the quartic  diagrams using the quartic vertices associated to the cubic Lagrangian reduce to the integral of  Green's function times the sum over $a$ and $n$ of~${\cal I}_3^{(n,a)}$.
%%%%%%%%%%%%%%%%%%%%%%%%%%%%%%%%%

\subsection{$CIS_{1PI}$ diagrams}

The $CIS_{1PI}$ diagrams read
\bea
CIS_{1PI}&=&-2\;{\rm Re}\int^t_{-\infty}dt_1\int^{t_1}_{-\infty} dt_2\langle\left[H_3(t_2),\left[\; H_3(t_1) ,\zeta_k(t)\right] \right]\zeta_k(t)\rangle=\\ \nonumber
&&2\;{\rm Re}\sum_{n,a} i \int^t_{-\infty}dt_1\int^{t_1}_{-\infty} dt_2\langle\left[H_3(t_2),\frac{\delta {\cal L}^{(n)}_3(t_1)}{\delta {\cal D}^{(n)}_a\zeta_a(t_1)}{\cal D}^{(n)}_aG_{\zeta_k}(t,t_1) \right]_k\zeta_k(t)\rangle= \\ \nonumber
&&2\;{\rm Re}\sum_{n,a}  \int^t_{-\infty}dt_1{\cal D}^{(n)}_a G_{\zeta_k}(t,t_1) \langle\left[i \int^{t_1}_{-\infty} dt_2 H_3(t_2),\frac{\delta {\cal L}^{(n)}_3(t_1)}{\delta {\cal D}_a^{(n)}\zeta_a(t_1)}\right]_k\zeta_k(t)\rangle\\ \nonumber
&&=\;\sum_{n,a}  \int^t_{-\infty}dt_1\,a(t_1)^{3+\delta}\, {\cal D}^{(n)}_a G^{(n)}_{\zeta_k}(t,t_1)\; {\cal I}_1^{(n,a)}(t_1) \ .
\eea
So the $CIS$ diagrams are the integral of the Green's function times the sum over $a$ and $n$ of ${\cal I}_1^{(n,a)}$.

\subsection{$CIM$ diagrams}
The $CIM$ diagrams read

\bea
CIM&=&-\langle\left[\int_{-\infty}^t dt_1\; H_3(t_1),\zeta_k(t)\right]\left[\int_{-\infty}^t dt_2\; H_3(t_2),\zeta_k(t)\right]\rangle\\ \nonumber 
&&=\sum_{n,m,a,b}\int^t_{-\infty}dt_1\;{\cal D}^{(n)}_a G^{(n)}_{\zeta_k}(t,t_1) \langle\left[\frac{\delta{\cal L}^{(n)}_3(t_1)}{\delta {\cal D}^{(n)}_a\zeta_a(t_1)}\right]_k \int_{-\infty}^t dt_2{\cal D}_b^{(m)} G_{\zeta_k}(t,t_2)\left[\frac{\delta{\cal L}^{(m)}_3(t_2)}{\delta {\cal D}_b^{(m)}\zeta_b(t_2)}\right]_k\rangle\\ \nonumber
&&=2\sum_{n,m,a,b}\int^t_{-\infty}dt_1\;{\cal D}^{(n)}_a G^{(n)}_{\zeta_k}(t,t_1) \langle\left[\frac{\delta{\cal L}^{(n)}_3(t_1)}{\delta {\cal D}^{(n)}_a\zeta_a(t_1)}\right]_k \int_{-\infty}^{t_1} dt_2{\cal D}_b^{(m)} G_{\zeta_k}(t,t_2)\left[\frac{\delta{\cal L}^{(m)}_3(t_2)}{\delta {\cal D}_b^{(m)}\zeta_b(t_2)}\right]_k\rangle\\ \nonumber
&&=\sum_{n,a}\int^t_{-\infty}dt_1\,a(t_1)^{3+\delta}\;{\cal D}^{(n)}_a G_{\zeta_k}(t,t_1)\; {\cal I}_2^{(n,a)}(t_1)
\eea
so the $CIM$ diagrams are the integral of the Green's function times the sum over $a$ and $n$ of ${\cal I}_2^{(n,a)}$.

\subsection{Quartic Diagrams from Cubic Lagrangian}

The fact that the cubic Lagrangian depends on $\dot\zeta$ means that the interaction picture quartic Hamiltonian receives a contribution that we call $H_{4,3^2}$, equal to
\be\label{eq:quartic-cubic}
H_{4,3^2}=\frac{1}{2}\frac{\delta P}{\delta\dot\zeta}\left(\frac{\delta \tilde {\cal L}_3}{\delta P}\right)^2=\frac{1}{2}\frac{\delta P}{\delta\dot\zeta}\sum_{b,n}{\cal D}^{(n,out)}_b\left(\frac{\delta \tilde{\cal L}_3}{\delta P}\right)\left(\frac{\delta {\cal L}^{(n)}_3}{\delta P_b}\right)\ ,
\ee
where in the second term we have explicitly stressed the sum over $b$ and we have integrated by parts any possible residual derivative (notice that the sign is re-absorbed in the definition of $\tilde {\cal L}_3$).
The resulting quartic diagram is
\bea
&&Quartic_{3,\d_t}=\\ \nonumber
&&2\; {\rm Re}\left\{\langle\left[i \int_{-\infty}^t dt_1 H_{4,3^2}(t_1),\zeta(t)\right]_k\zeta_k(t)\rangle\right\}= {\rm Re}\left\{\langle\left[i \int_\infty^t dt_1\frac{\delta P}{\delta\dot\zeta}\left(\frac{\delta\tilde {\cal L}_3}{\delta P}\right)^2,\zeta(t)\right]_k\zeta_k(t)\rangle\right\} \\ \nonumber
&&=- {\rm Re}\sum_{n,a}\left\{ \int_{-\infty}^t dt_1 \;{\cal D}^{(n)}_a G_{\zeta_k}(t,t_1)\ \times\right. \\  \nonumber
&&\left. \langle\left[\left(\frac{\delta^2 {\cal L}^{(n)}_3(t_1)}{\delta {\cal D}^{(n)}_a\zeta_a\delta \dot\zeta_b}\right) {\cal D}_b^{(n,out)} \left(\frac{\delta\tilde {\cal L}_3(t_1)}{\delta P}\right)+ {\cal D}_b^{(n,out)}\left(\frac{\delta \tilde{\cal L}_3(t_1)}{\delta P}\right)\left(\frac{\delta^2 {\cal L}^{(n)}_3(t_1)}{\delta {\cal D}^{(n)}_a\zeta_a\delta \dot\zeta_b}\right)\right]_k\zeta_k(t)\rangle\right\}\\\nonumber
&&= \sum_{n,a}\int^t_{-\infty}dt_1\,a(t_1)^{3+\delta}\;{\cal D}_a^{(n)} G_{\zeta_k}(t,t_1)\; {\cal I}_3^{(n,a)}(t_1)\ .
\eea
So the $Quartic_{3,\d_t}$ diagrams are the integral of the Green's function times the sum over $a$ and $n$ of ${\cal I}_3^{(n,a)}$.
By summing the final expressions from the $CIS_{1PI}$, $CIM$ and $Quartic_{3,\d_t}$, we obtain the remarkably simple formula in~eq.~(\ref{eq:master}), as we wanted to show.

\section{Time-(in)dependence of $\zeta$ from cubic diagrams\label{sec:cubic}}

We can now ask ourselves if the contribution from the diagrams considered in the former section can lead to a time dependence on the $\zeta_k$ correlation function after the comoving mode $k$ has crossed the horizon. 

\subsection{Quartic$_{\d_i}$ diagrams}

To understand wether the diagrams considered so far can lead to a time dependence, it will turn out to be useful to first add the quartic diagrams that are associated to the rescaling of the spatial derivatives in the cubic vertices. We call these Quartic$_{\d_i}$. They take the form
\be\label{quartic_spatial}
Quartic_{\d_i}=-\sum_n \d_i\d_t^n\zeta\int d\zeta \frac{\d {\cal L}_3}{\d(\d_i\d_t^n\zeta)}\ ,
\ee
The symbol $\int d\zeta$ represents the fact that we multiply $\d {\cal L}_3/\d(\d_i\d_t^n\zeta)$ by $\zeta$ if there is no $\zeta$ without any derivative acting on it in $\d {\cal L}_3/\d(\d_i\d_t^n\zeta)$, we multiply by $\zeta/2$ if there is one $\zeta$ without any derivative acting on it~\footnote{The last remaining option, two $\zeta$'s in $\d {\cal L}_3/\d(\d_i\d_t^n\zeta)$ without any derivatives acting on them, is forbidden by rotational invariance.}. The reason why we wish to include these quartic diagrams with the former is due to the fact that whenever an operator contains a spatial derivative, we expect that in the presence of long $\zeta$ mode the coordinates are effectively rescaled in a form $\d_i\rightarrow e^{-\zeta}\d_i$. As we will explain more in detail, the former interactions do not take into account of this rescaling, which is instead implemented by the quartic terms we are singling out. More formally, we can understand the presence of these terms in the following way. In the ADM parametrization
\be
ds^2=- N^2 dt^2+h_{ij} (dx^i+N^i dt)(dx^j+N^j dt)\ ,
\ee
$\zeta$ gauge and the $\zeta$ perturbation are defined by fixing the spatial diff.s by imposing the spatial metric to take the form
\be
h_{ij}= a(t)^2e^{2\zeta(\vec x,t)} \delta_{ij}\ ,
\ee
and the time diff.s are fixed by imposing the inflaton perturbations to be zero.
% {\bf ?!?!?!?!?!?!?!?!?! attention to this area: quite delicate ?!?!?!?!?!?!?!!?!?!?!?!?!?!?!?!Question: is there a zero-mode inflaton perturbation ???!!?!? !?!}.  
This gauge choice leaves some zero-mode spatial diff.s unfixed. For example those that are associated to a time dependent rescaling and translation of the spatial coordinates:
\be\label{eq:diff1}
x^i\quad\rightarrow \quad x^i=e^{\beta(t)}\tilde x^i+C^i(t)\ ,
\ee 
with $\beta(t),\; C_i(t)$ generic functions of time. Under this rescaling, $\zeta$ and $N^i$ transform as
\bea\label{eq:diff2}
&&\zeta\quad\rightarrow\quad\tilde\zeta=\zeta+\beta(t)\ ,\\ \nonumber
&&N^i\quad\rightarrow \quad \tilde N^i=N^i e^{-\beta}+ \dot\beta(t) \tilde x^i+e^{-\beta}\dot C^i(t).
\eea
Thus the $\zeta$ zero mode has not been gauge fixed. For our purposes, we therefore learn that the $\zeta$ action must be diff. invariant under this restricted group of diff.s. Therefore, any combination of $\d_i$ must actually take the form $e^{-\zeta} \d_i$ to be diff. invariant. By Taylor expanding this exponential, we clearly see that there is a connection between linear and quadratic terms, or from cubic and quartic terms.

To be even more explicit,  let us give some examples. Given a vertex in the cubic Lagrangian, we identify the necessary vertex to be considered from the quartic Lagrangian in the following way
\bea\label{eq:example}
{\cal L}_3 \quad\supset \quad \zeta(\d_i\zeta)^2 \qquad&\rightarrow& \qquad  {\cal L}_4 \quad\supset\quad - \zeta^2(\d_i\zeta)^2\ , \\ \nonumber
{\cal L}_3 \quad\supset \quad \dot\zeta(\d_i\zeta)^2 \qquad&\rightarrow& \qquad {\cal L}_4 \quad \supset\quad-  2\zeta\dot\zeta(\d_i\zeta)^2\ .
\eea

\subsection{Time independence and the consistency condition}

It is useful to split formula $(\ref{eq:master})$ into the sum of two terms. Let us introduce a time $t_{k_{out}}$ quite after the mode $k$ has crossed the horizon $k/a(t_{k_{out}})=\epsilon_{out} H(t_{k_{out}})$, with $\epsilon_{out}\ll1$. Eq.~(\ref{eq:master}) can be written as
\bea\label{eq:master2}
&&\!\!\!\!\!\!\!\!\!\!\langle\zeta_k\zeta_{k}\rangle_{CIS_{1PI}+CIM+Quartic_{3,\d_t}}=\lim_{\epsilon\rightarrow 0}\left(\int_{-\infty}^{t_{k_{out}}} dt_1+\int_{t_{k_{out}}}^{t} dt_1\right)\\ \nonumber 
&& \left[ a(t_1)^{3+\delta}\sum_{a,n}{\cal D}_a^{(n)}G_{\zeta_k}(t,t_1)2{\rm Re}\langle\left[\frac{1}{a(t_1)^{3+\delta}}\frac{\delta{\cal L}_3^{(n)}(t_1)}{\delta {\cal D}^{(n)}_a\zeta_a(t_1)}\right]_kU_{int}^\dag(t_1,-\infty)\zeta_{k,I}(t)U_{int}(t_1,-\infty)\rangle \; e^{\epsilon k t_1}\right] \ . 
\eea
The contribution from the first term represents the case where the three-point function is evaluated at a time before the time $t_{k_{out}}$, while the second integral represents the contribution from evaluating the contribution of the three-point function from time $t_{k_{out}}$ up to the present time $t$.

Clearly, the first term is time-independent. The only dependence on $t$ appears in the last term $\zeta_I(t)$, the interaction picture field that is constant at $t\gg t_{k_{out}}$. Let us therefore concentrate on the second term. Since we are considering times when the mode $k$ is very outside of the horizon, we can expand the Green's function at late times $k/a(t_1)\ll H$. In conformal time, we have
\be
G_{\zeta_k}(\eta,\eta_1)\simeq\frac{H^2}{3}\left(\eta^3-\eta_1^3\right)\theta(\eta-\eta_1)\ ,
\ee
obtaining
\bea\label{eq:master3}\nonumber
&&\langle\zeta_k\zeta_{k}\rangle_{CIS_{int}+CIM+Quartic_{3,\d_t},t}\simeq\lim_{\epsilon\rightarrow 0}\int_{\eta_{k_{out}}}^{\eta} d\eta_1\; \left(-\frac{1}{H\eta_1}\right)^{4+\delta}\sum_{a,n}{\cal D}_a^{(n,out)}\frac{H^2}{3}\left(\eta^3-\eta_1^3\right)\theta(\eta-\eta_1) \\ 
&&\qquad\qquad\qquad\qquad2\,{\rm Re}\langle\left[\frac{1}{a(\eta_1)^{3+\delta}}\frac{\delta{\cal L}_3^{(n)}(\eta_1)}{\delta {\cal D}^{(n)}_a\zeta_a(\eta_1)}\right]_kU^\dag(\eta_1,-\infty)\zeta_{k,I}(\eta)U(\eta_1,-\infty)\rangle \; e^{\epsilon\, k\log(a(\eta_1))/H} \nonumber\, ,  \\
\eea
where the subscript $_t$ in $\langle\zeta_k\zeta_{k}\rangle_{CIS_{1PI}+CIM+Quartic_{3,\d_t},t}$ refers to the fact that we are concentrating only on the time dependent part, and where the appearance of ${\cal D}_a^{(n,out)}$ is due to the fact that the commutators of $[\zeta_k,\dot\zeta_k]$ and $[\zeta_k,\zeta_k]$ scale  in the same way at late times. Neglecting any possible time dependence from the terms in the second line, we see that naively 
the time integral diverges as 
\be
\int^\eta d\eta_1\frac{1}{\eta_1^{1+\delta}}\sim\frac{1}{\eta_1^\delta}\quad\rightarrow\quad \log(-\eta)\sim H t \quad{\rm as}\quad {\delta\rightarrow 0}\ .
\ee
We see the potential risk of linear infrared divergencies in cosmic time $t$  (logarithmic in conformal time $\eta$) in the case the three-point function's contribution, that we have neglected in this formula, does not decay in time. Contributions from terms with ${\cal D}_a^{(n,out)}$ being non-unity are clearly more convergent by powers of $\eta_1$.

Let us therefore concentrate on the three point function, which can be schematically written as a convolution:
\be
\sim\int d^{3+\delta}q\; \langle{\cal D}_1\zeta_{\vec q}(t_1){\cal D}_2\zeta_{\vec k-\vec q}(t_1)\; U(t_1,-\infty)^\dag\zeta_{k,I}(t)U(t_1,-\infty)\rangle
\ee
where ${\cal D}_{1,2}$ represent generic differential operators (including the identity operator) that could be present. The integral in $q$ runs from very small wavenumbers (much smaller than $k$) up to infinity because we are working in dimensional regularization. 

The contribution from momenta smaller than $k/\epsilon_{out}$ cannot give a time dependence. This is so because as these modes are longer than $k/\epsilon_{out}$, the three-point function is evaluated when all the Fourier modes are very outside of the horizon. A remarkable property of the cubic interaction Lagrangian of $\zeta$, which can be traced back to the original diff. invariance of the Lagrangian, is the fact that it can be written in a form where there are {\it no} operators with either no derivative or just a time derivative~\cite{Maldacena:2002vr}. This means that if we decide to consider the contribution from terms where ${\cal D}_a^{(n,out)}$ is absent, so that they are potentially IR divergent, we are forced to consider an operator $\left[\delta{\cal L}_3^{(n)}(\eta_1)/\delta ({\cal D}^{(n)}_a\zeta_a(\eta_1))\right]_k\sim\zeta_{\vec q}(t_1)\zeta_{\vec k-\vec q}(t_1)$ with at least a derivative acting on one of the two operators. This therefore leads to a time-convergent integral~\footnote{Similar conclusion applies also to the case where we consider the operator 
\be
\frac{\d_i}{\d^2}\dot\zeta\d_i\zeta\dot\zeta\ .
\ee
as even if we remove first non-local term by inserting it in the Green's function, we are left with $\d_i\zeta\dot\zeta$ that has enough derivatives to compensate for the non local term.}. 

We are finally lead to consider the remaining  part of the integral where we include modes $q\gtrsim k/\epsilon_{out}$. These modes are at horizon crossing or well inside the horizon when the three-point function is evaluated, and so, contrary to what happens in the former regime $q\lesssim k/\epsilon_{out}$, there is no suppression for derivatives acting on these modes. However, in this regime we can use a remarkable property of the three-point function in the regime $k\ll q$, the so called `consistency condition' of the three-point function, which states that at leading order in $k/q\ll1$, $k/(a H)\ll1$, the three-point function has the following form
\bea\label{eq:consistency}
&&\langle\left[\frac{1}{a(\eta_1)^{3+\delta}}\frac{\delta{\cal L}_3^{(n)}(\eta_1)}{\delta {\cal D}^{(n)}_a\zeta_a(\eta_1)}\right]_{k,\;(q\gg k)}\zeta_{k,I}(\eta_1)\rangle\simeq \\ \nonumber
&&\qquad\quad \quad \simeq\frac{1}{q^{3+\delta}}
\frac{\partial \langle \left[q^{3+\delta}\frac{1}{a(\eta_1)^{3+\delta}}\frac{\delta{\cal L}_3^{(n)}(\eta_1)}{\delta {\cal D}^{(n)}_a\zeta_a(\eta_1)}\right]_q\rangle}{\partial\log q} \langle\zeta_k(\eta_1)^2\rangle+{\cal{O}}\left({\rm Max}\left[\left(\frac{k}{a(\eta_1) H(\eta_1)}\right)^2,\left(\frac{k}{q}\right)^2\right]\right)\ .
\eea 
The last term represents the subleading correction to the squeezed limit.  Let us understand the ${\rm Max}\left[\frac{k}{a(\eta_1) H(\eta_1)},\frac{k}{q}\right]$ term. If we expand in gradients in the long wavelength fluctuation, the natural quantity to consider is clearly the physical wavenumber $k/(a H)$.  So, this is the natural size of the correction in the squeezed limit. The calculation of the three-point function in this limit involves a time integral in a variable that we can call $\eta_2$. Subleading corrections in the squeezed limit are contained in the integrand are proportional to $k/(a(\eta_2) H(\eta_2))$. If the short modes $q$ are longer than the horizon at the time $\eta_1$, then the time integral is peaked at the time $\eta_2 $ when the modes $q$ crossed the horizon. This gives $q/(a(\eta_2) H(\eta_2))\sim 1$, which gives a correction of the form $k/q$. If the modes $q$ are instead still inside the horizon at $\eta_1$, the integral is peaked at $\eta_2\sim \eta_1$, giving a correction of the form $k/(a(\eta_1)H(\eta_1))$.

There are two subtleties to discuss about the above formula (\ref{eq:consistency}). The first regards the case in which the operator $\left[\delta{\cal L}_3^{(n)}(\eta_1)/\delta ({\cal D}^{(n)}_a\zeta_a(\eta_1))\right]_k$ contains spatial derivatives of the short modes, for example if it is of the form $(\d_i\zeta)^2$. In this case the consistency condition does not hold. The consistency condition implies  that in the squeezed limit the 3-point function follows directly from the fact that in this limit the long mode acts as a rescaling of the spatial coordinates $\vec x\rightarrow e^{-\zeta}\vec x$. However, when we compute the 3-point function with the usual formulas $\sim [\int dt\,H_{int},\zeta^3]$, we are evolving in the interaction picture the operators $\zeta$, not the spatial coordinates themselves. This means that evaluation of the 3-point function amounts to effectively rescaling the {\it argument} of the operators $\zeta(\vec x)\rightarrow\zeta(e^{-\zeta}\vec x)$. The computation does not implement the rescaling of the spatial derivatives, simply because they `go along with the ride', unaffected by the interacting Hamiltonian. Formula (\ref{eq:consistency}) does not hold. Although this seems to challenge the very intuitive result that a long wavelength $\zeta$ acts as a rescaling of the coordinates,  diff. invariance provides a solution. The quartic vertex Quartic$_{\d_i}$in eq.~(\ref{quartic_spatial}) provides precisely the contact term necessary to rescale the coordinates in the spatial derivative. So, eq.~(\ref{eq:consistency}) holds after we add to all the diagrams considered so far also the Quartic$_{\d_i}$. In  App.~\ref{app:consistency-inside}, we discuss examples of three-point functions in the squeezed limit in which one of the modes has much longer wavelength than the others, involving short modes that are still inside the horizon and that are acted upon by space and time derivatives. There we show that the consistency condition holds after the addition of the relevant contact operators.

The second subtlety in using (\ref{eq:consistency}) is that in the three-point function we are computing the last term should be
$$
U_{int}(\eta_1,-\infty)^\dag\zeta_{k,I}(\eta)U_{int}(\eta_1,-\infty)\ ,
$$ 
which is different from $\zeta_{k,I}(\eta_1)$. This is equivalent to the situation where we were to arbitrarily shut down $H_{int}$ at $t_1$ and the theory become free after that. Even though this is not the case in the actual physical system, it can be straightforwardly realized that this difference does not matter, because at the time $t_1$ the $k$-mode is already well outside the horizon. We therefore are free to use (\ref{eq:consistency}) at leading order in $k/(a H)$. 

By substituting (\ref{eq:consistency}) into (\ref{eq:master3}), we obtain:
\bea\label{eq:master4}
&&\langle\zeta_k\zeta_{k}\rangle_{CIS_{1PI}+CIM+Quartic_{3,\d_t}+Quartic_{\d_i},t}\simeq\lim_{\epsilon\rightarrow 0}\int_{\eta_{k_{out}}}^{\eta} d\eta_1\; \left(-\frac{1}{H\eta_1}\right)^{4+\delta} \\ \nonumber
&&\quad\sum_{a,n}{\cal D}_a^{(n,out)}\frac{H^2}{3}\left(\eta^3-\eta_1^3\right)\theta(\eta-\eta_1)  2\,{\rm Re}\int^{+\infty}_{k/\epsilon_{out}} d^{3+\delta}q\;\frac{1}{q^{3+\delta}}\frac{\partial \langle \left[q^{3+\delta}\frac{1}{a(\eta_1)^{3+\delta}}\frac{\delta{\cal L}_3^{(n)}(\eta_1)}{\delta {\cal D}^{(n)}_a\zeta_a(\eta_1)}\right]_q\rangle}{\partial\log q} \langle\zeta_k(t)^2\rangle\;
\ . 
\eea
%{\bf ?!?!?!?!?!?!?!?!?!? I have removed the part on the odd spatial indexes in the 3-pt function, as I think it is irrelevant?!?!?!?!?!?!?!!?!?!?!?}
The rotational integral is trivially performed, and the remaining momentum $q$-integral is a total derivative. 
%{\bf Comment on the contribution fron infinty ?!??!?!?!?!?!!?!?!?!?!?!?!?!?!?!?!?it is zero as the integral is convergent ?!}
This leads to
\bea\label{eq:master5}\nonumber
\langle\zeta_k\zeta_{k}\rangle_{CIS+CIM+Quartic_{3,\d_t}+Quartic_{\d_i},t}&\simeq&\lim_{\epsilon\rightarrow 0}\int_{\eta_{k_{out}}}^{\eta} d\eta_1\; \left(-\frac{1}{H\eta_1}\right)^{4+\delta}\sum_{a,n}{\cal D}_a^{(n,out)}\frac{H^2}{3}\left(\eta^3-\eta_1^3\right)\theta(\eta-\eta_1) \\ 
&&8\pi\,\left. \langle \left[\frac{1}{a(\eta_1)^{3+\delta}}\frac{\delta{\cal L}_3^{(n)}(\eta_1)}{\delta {\cal D}^{(n)}_a\zeta_a(\eta_1)}\right]_q\rangle \right|_{q=k/\epsilon_{out}} \langle\zeta_k(t)^2\rangle\;
\ ,
\eea
where the contribution from $q=\infty$ is zero as the integral is made convergent in dim-reg. As we evaluate the term 
\be
\left.\langle \left[q^{3+\delta}\frac{1}{a(\eta_1)^{3+\delta}}\frac{\delta{\cal L}_3^{(n)}(\eta_1)}{\delta {\cal D}^{(n)}_a\zeta_a(\eta_1)}\right]_q\rangle\right|_{q=k/\epsilon_{out}} 
\ee
and we take the limit $\eta_1\rightarrow 0$ as $\eta\rightarrow 0$, we notice the property of the cubic $\zeta$-Lagrangian that we mentioned before: there is {\it no} operator $\left[\delta{\cal L}_3^{(n)}(\eta_1)/\delta ({\cal D}^{(n)}_a\zeta_a(\eta_1))\right]_k$ that does not vanish as some power of $k/(a(\eta_1)H(\eta_1))\sim k\eta_1\rightarrow 0$. This is so because  in order for this term to have any chance to contribute at late times we had to restrict ourselves to choosing an operator that had at least one derivative acting on one of the two $\zeta$ operators. Since this terms is evaluated when momenta are outside the horizon, it vanishes as $\eta_1\rightarrow 0$. This means that the resulting time integral is convergent. 

We stress that there is no time dependence because, as a result of the consistency condition, the integrand in the internal momenta $q$ becomes a total derivative. If this had not been the case, it would have been less trivial to show that the result of the integration leads to a time independent answer~\footnote{Let us comment on the contribution of the subleading corrections in eq.~(\ref{eq:consistency}), which do not take the form of  a total derivative. Those contributions are not scale invariant in the external wavenumber $k$, having one additional factor of $k$ in the numerator with respect to the leading, scale invariant contribution. This means that the resulting contribution goes to zero at late times as $(k\eta_1)^2$ and so they lead to a time-convergent contribution as $\eta\rightarrow 0$. The fact that the contribution to the subleading corrections in eq.~(\ref{eq:consistency}) is not scale invariant comes from the following. If we consider the contribution from any fixed momentum shell in $q$ between $q\sim k/\epsilon_{out}$ to $q\sim \gamma k$ with $\gamma\gg 1/\epsilon_{out}$, with $\gamma$ a time independent number small enough so that $q$ is outside the horizon, the contribution from that shell of momenta goes to zero as some power of $k\eta_1$.  This is so because the operator $\left[\delta{\cal L}_3^{(n)}(\eta_1)/\delta ({\cal D}^{(n)}_a\zeta_a(\eta_1))\right]_k$ contains some derivatives of the fields. This means that the contributions in the integrand coming from momenta outside of the horizon is peaked at those momenta at horizon crossing $q\sim a H$, which meanS that the subleading corrections are of the form $k/q\sim k/(a H)$ and so the integrand goes to zero as $\eta_1\rightarrow 0$. Finally, the contribution from momenta $q$ that are inside the horizon is explicitly down by powers of $k/(a H)$ and so they are as well convergent.}.
 
This result can probably  be stated more intuitively by simply noticing that the consistency condition implies that in the extreme squeezed limit $k\ll q$ the effect of the long mode on the dynamics is to do {\it nothing}: its effect is simply a trivial rescaling of the comoving momenta. Since we compute the integrals over the whole high momentum modes, this rescaling has no effect apart for changing the boundary of integration for the most infrared modes of order $k/\epsilon_{out}$. But the integral has no support in that region. This is a simple explanation of the reason why the loop integral becomes a total derivative in the squeezed limit.

This is enough to make the subleading corrections time convergent. We have at this point gone through the whole phase space in $CIM+CIS+Quartic_{3,\d_t}+Quartic_{\d_i}$ diagrams, finding that their sum leads to no time dependence.

\vspace{0.5cm} 
{\bf A note on the counterterms:} It is important to realize that  (\ref{eq:master5}) is the result of the full loops integrals in the squeezed limit $k\ll q,\; k/(aH)\ll1$. The integral is therefore UV finite, even in the limit in which we send the number of spatial dimensions to three, or the regulator to infinity. This is a very important consistency check. If the integral in this regime were to be UV divergent we would have had a divergent time dependence piece and we would have needed a counterterm that cancelled the divergent time-dependence of~$\zeta$. But there are no counterterms in the action that induce a time-dependence for $\zeta$ because that is equivalent to inducing at quadratic level a mass for $\zeta$ which does not happen for the terms allowed by the symmetries. As we will see in the next section, the only quadratic counterterms that induce a mass for $\zeta$ are the ones associated to the tadpole terms, that induce {\it also} a linear tadpole for $\zeta$. We will verify they will exactly cancel the time-dependence from the diagrams built with the quartic vertices. 

\subsection{Example\label{sec:example1}}

It is instructive to find a simple example where this can be seen explicitly. Thanks to the Effective Field Theory of Inflation~\cite{Cheung:2007st}, it is possible to find a consistent inflationary Lagrangian which has the properties we discussed~\footnote{The Effective Field Theory of Inflation is a quite powerful new formalism to describe the theory of inflation in very general terms. A sample of recent works that have been developing it is given by~\cite{Cheung:2007st, Cheung:2007sv,Senatore:2009cf,eft} }. By parametrizing the fluctuations in terms of the Goldstone boson $\pi$ and going to the decoupling limit, the algebra becomes very simple. Let us take for example the following Lagrangian in the decoupling limit
\be\label{eq:efttoyact}
S=\int d^4x\,\sqrt{-g}\;\left[-\dot H\mpl^2\left(\dot\pi^2-\frac{1}{a^2}(\d_i\pi)\right)+M^4(t+\pi)\left(\dot\pi^2+\ldots\right)\right]
\ee
where $\ldots$ represent cubic or quartic terms in $\pi$ that have one derivative acting on each fluctuation. Those terms do not lead to any diagram with an explicit time dependence, and we neglect them here. For illustrative purposes, let us suppose now that the function $M^4(t)$ depends linearly on time. By Taylor expanding in $\pi$, we notice that we have the cubic interaction
\be
{\cal L}_3= \d_t\left( M^4(t)\right) \pi\dot\pi^2\ .
\ee
This interaction in very dangerous. If we imagine forming a loop with two of these vertices and using a $\pi$ in the first vertex to contract with the external leg, the resulting diagram will become time-dependent. This means that time-independence can come only from a quartic interaction. Indeed, this is exactly the kind of cubic Lagrangians that leads to a non-trivial ${\cal H}_{4,3}$.

Bu concentrating only on the effects proportional to $(\d_t M^4)^2$, the action can be recast as
\be
{\cal S}=\int d^4x\,\sqrt{-g}\;\left[\frac{-\dot H\mpl^2}{c_s^2}\left(\dot\pi^2-\frac{c_s^2}{a^2}(\d_i\pi)^2\right)+\left( \d_t\left( M^4(t)\right)  \pi\dot\pi^2\right)\right]\ .
\ee
%where we have neglected all the interactions that do not lead to a time-dependence in the diagrams. 
The speed of sound is $c_s^2=-\dot H\mpl^2/(-\dot H\mpl^2+M^4(t))$. The momentum conjugate to~$\pi$,~$P$, is given by
\be
P=\frac{\delta {\cal L}}{\delta \dot\pi}=2a^3\left(\frac{-\dot H\mpl^2}{c_s^2}+ \d_t\left( M^4(t)\right)  \pi\right)\dot\pi\ ,
\ee 
and the Hamiltonian is therefore
\be
{\cal H}=P\,\dot\pi(\pi,P)-{\cal L}(\pi,\dot\pi(\pi,P))=\frac{P^2}{4 a^3 \left(\frac{-\dot H\mpl^2}{c_s^2}+ \d_t\left( M^4(t)\right)  \pi\right)}+a^3 \left(-\dot H\mpl^2\right)\frac{1}{a^2}(\d_i\pi)^2\ .
\ee
We can identify the quartic Hamiltonian of order $(\d_t M^4)^2$ to be
\be
{\cal H}_{4,3}=\frac{ \left[\d_t\left( M^4(t)\right)\right]^2}{4 a^3 \left(\frac{-\dot H\mpl^2}{c_s^2}\right)^3} P^2\pi^2=a^3\frac{ \left[\d_t\left( M^4(t)\right)\right]^2}{ \frac{-\dot H\mpl^2}{c_s^2}} \dot\pi_I^2\pi^2_I\ .
\ee
where in the second passage we have written the expression in terms of the interaction picture fields. It can be easily checked that this agrees with (\ref{eq:quartic-cubic}).
Quartic diagrams built with ${\cal H}_{4,3}$ lead also to time dependence, a time-dependence that indeed cancels the one from the cubic diagrams built with $(\d_t(M^4(t))\pi\dot\pi^2$. This example is discussed in detail in appendix \ref{timederiv}.

\section{Time-(in)dependence of $\zeta$ from quartic diagrams}

In order to complete the study of the possible infrared effects we need to look at the contribution from the remaining quartic interactions $H_4\supset H_{4,4}=-{\cal L}_4-{\cal L}_{4,\d_i}$, where ${\cal L}_{4,\d_i}$ represents the terms that were borrowed in the former section to give the $Quartic_{\d_i}$ diagrams. These remaining diagrams contribute to the two point function  as
\bea\label{eq:masterquad}
&&\langle\zeta_k\zeta_{k}\rangle_{Quartic_4}=\\ \nonumber 
&&-\lim_{\epsilon\rightarrow 0}\int_{-\infty}^t dt_1\; a(t)^{3+\delta}\sum_{a,n}{\cal D}_a^{(n)}G_{\zeta_k}(t,t_1)2{\rm Re}\langle\left[\frac{1}{a(t_1)^{3+\delta}}\frac{\delta{\cal L}_4^{(n)}(t_1)}{\delta {\cal D}^{(n)}_a\zeta_a(t_1)}\right]_{k}\zeta_{k}(t_1)\rangle \; e^{\epsilon k t_1}\ .
\eea

Like in the former section, it is straightforward to see that the factor before the four-point function on the left of the above formula leads to a  time dependence proportional to $H t$ if the four-point function does not have a suppression at late time. Contrary to what happened in the former section with the three-point function after it was integrated over comoving monenta, there is no such a cancellation from diagrams within $H_4$. So there is a subset of diagrams that naively lead to a time-dependence. We are now going to show that there is a cancellation that leads to absence of  a time dependence of $\zeta$ at late times after adding a new set of diagrams. These new diagrams come from effectively quartic vertices that arise when we insert the couterterms for the tadpoles.

Let us see this in detail. At one loop order the first diagrams we should consider are the tadpole diagrams, that can be written as
\bea\label{eq:mastertad}
\langle\zeta_k\rangle_{Tad}=\lim_{\epsilon\rightarrow 0}\int_{-\infty}^t dt_1\; a(t)^{3+\delta}\sum_{a,n}{\cal D}_a^{(n)}G_{\zeta_k}(t,t_1)\langle\left[\frac{1}{a(t_1)^{3+\delta}}\frac{\delta{\cal L}_3^{(n)}(t_1)}{\delta {\cal D}^{(n)}_a\zeta_a(t_1)}\right]\rangle \; e^{\epsilon k t_1}\ .
\eea
Very simple counting arguments shows that these diagrams can lead to a time dependence of the zero mode $\zeta_{k=0}$. If these diagrams are not zero it is because we are expanding around the wrong unperturbed history. Indeed, by translation invariance, only the $k=0$ mode is directly affected, and the zero mode can be totally reabsorbed in the definition of the unperturbed history.  However this does not mean that these diagrams affect only the zero mode: they can be attached with a cubic vertex to a propagator to affect the two point function of modes at finite $k$ in a non-1PI diagram (Fig.~\ref{CIS_non_1PI}), and possibly induce a time dependence even there. The fact that this diagrams is not zero is clearly a nuisance.

Fortunately, these diagrams can be set to zero by inserting proper counterterms. In order to cancel tadpole diagrams, they must start linear in the fluctuations. In principle, there are many possible operators of this form, but luckily we can use a theorem proved in the context of the Effective Field Theory of inflation~\cite{Cheung:2007st}. It states that all the possible tadpole counterterms can be reduced to just two operators~\footnote{We stress that this is one of the advantages of using the Effective Field Theory of Inflation: by concentrating directly on the fluctuations, it allows immediately to identify the operators with the correct number of fluctuating fields to be tadpole counterterms.}. In unitary gauge, these are 
\be
{\cal S}_{tad,counter}=\int d^4x\sqrt{-g}\;\left[g^{00} \delta M^4(t)+\delta\Lambda(t)\right]\ .
\ee
Up to one loop level, the terms starting linear in the fluctuations take the form
\be\label{eq:Stadpole}
{\cal S}_{tad}=\int d^4x\sqrt{-g}\;\left[ g^{00} \left(\mpl^2\dot H+\delta M^4\right)-  \mpl^2\left(\left(3H^2+\dot H\right)+\delta\Lambda\right)\right]\ , 
\ee
The coefficients $\dot H\mpl^2$ and $-\mpl^2(3H^2+\dot H)$ are uniquely fixed by the background, as proven in~\cite{Cheung:2007st}, while the terms $\delta M^4$ and $\delta\Lambda$ represent the one-loop counterterms that are chosen to cancel the tadpole diagrams. The most important point that we need to realize is that these operators that start linear in the fluctuations necessarily contain higher order terms. This is so because of the non-linear realization of time diffs. In particular this means that there will be quadratic terms that can contribute to the two-point function effectively as one-loop terms. In this section we are going to prove that they exactly cancel the quartic diagrams constructed with $H_4$ that would lead to a time dependence.

\subsection{Example:}

\begin{figure}[h]
\begin{center}
\includegraphics[width=14cm]{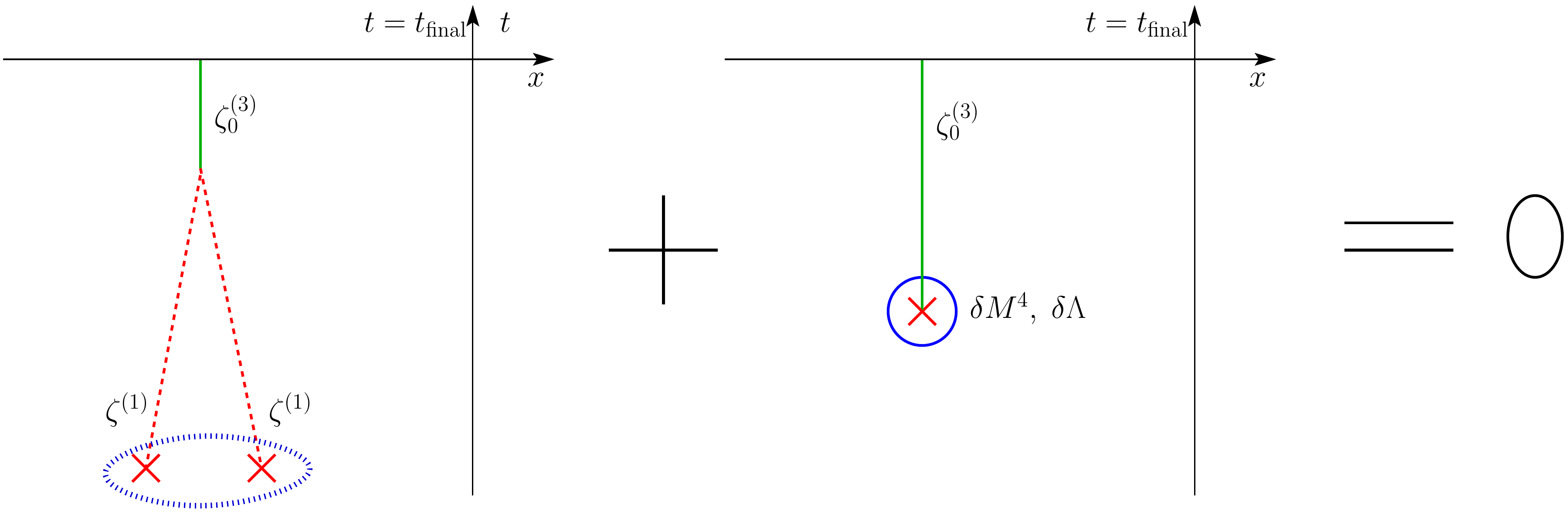}
\caption[]{ Cancellation between the tadpole diagram and the tadpole counterterm.}
\label{tadpole_cancellation}
\end{center}
\end{figure}

\begin{figure}[h]
\begin{center}
\includegraphics[width=17cm]{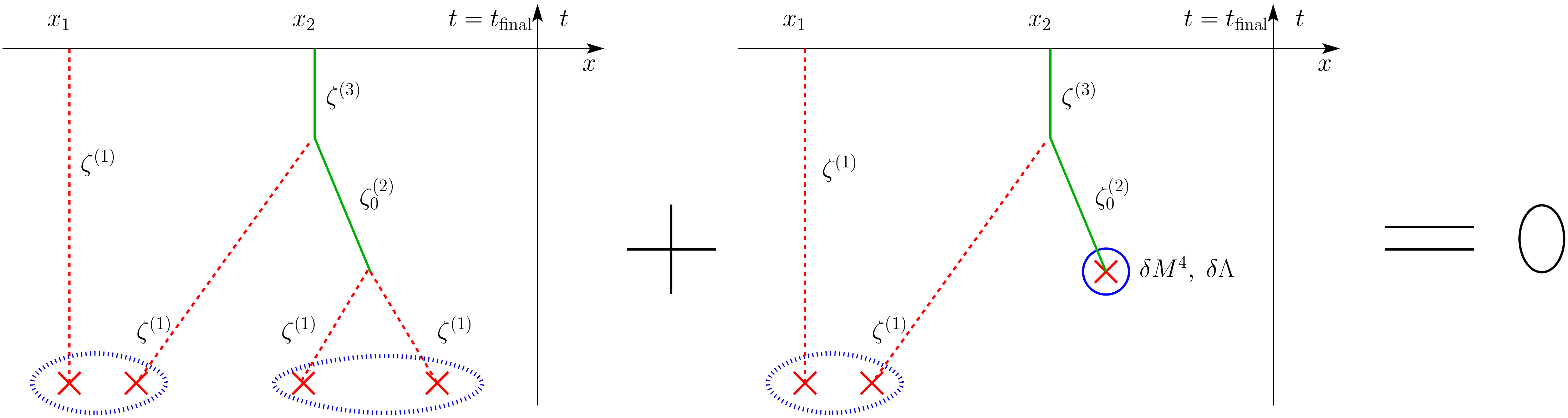}
\caption[]{ Cancellation of the $CIS_{non-1PI}$ diagrams with the $CIS_{non-1PI}$ diagrams constructed with the tadpole counterterms }
\label{tadpole_2_point_cancellation}
\end{center}
\end{figure}

Since the algebra quickly becomes  very complicated, we use the Effective Field Theory of Inflation~\cite{Cheung:2007st} to find a consistent inflationary model where this cancellation can be studied in the simplest context. Let us consider the following Lagrangian in unitary gauge
\be
{\cal S}=\int d^4x\sqrt{-g}\;\left[ g^{00} \left(\mpl^2\dot H+\delta M^4\right)-  \mpl^2\left(\left(3H^2+\dot H\right)+\delta\Lambda\right)+M_3^4\left(\delta g^{00}\right)^3\right]
\ee
and let us imagine that $M_3^4$ depends rapidly linearly in time. This means that we can concentrate on that interaction and study it in the decoupling limit. Upon reinserting the Goldstone boson $\pi$ by performing a time-diff $t\rightarrow t+\pi$, the Lagrangian reduces to
\bea \nonumber
&&{\cal S}=\int d^4x\sqrt{-g}\;\left[(-1-\dot\pi+(\d\pi)^2)\left(\mpl^2\dot H+\delta M^4(t+\pi)\right)-  \mpl^2\left(\left(3 H^2+\dot H\right)+\delta\Lambda\right)+\right.\\
&&\qquad\qquad\qquad\left.M_3^4(t+\pi)\dot\pi^3\right]\ ,
\eea
where we have stopped at quartic level and we have kept only the interactions proportional to $M_3(t)$. By Taylor expanding the last term, we have a vertex of the form $\dot{M_3^4}\pi\dot\pi^3$ which, if we contract $\dot\pi$ as the final leg in the Green's function, leads to a quartic diagram that naively induces a time-dependence. Let us see how it cancels with the operators induced by the tadpole counterterms.
By the non-linear realization of time diffs., this same operator starts cubic, and it therefore induces a tadpole. All diagrams with only one vertex can be most simply studied directly in the Lagrangian by taking the expectation value of the quadratic operators contracted in the loop, and studying the resulting quadratic Lagrangian. This is equivalent to resuming all the non-1PI diagrams obtained by multiple insertion of the same loop. So we notice that the last term induces a tadpole term of the form
\be
\delta{\cal S}_{3\rightarrow 1}=\int d^4x\sqrt{-g}\left[3 M_3^4(t)\delta g^{00}\langle(\delta g^{00}){}^2\rangle\right]\ .
\ee
This means that in order to cancel this diagram we have to choose $\delta M^4$ as
\be
\delta M(t)^4=-3 M_3^4(t)\langle(\delta g^{00}){}^2\rangle\ .
\ee
This is shown diagrammatically in Fig.~\ref{tadpole_cancellation} where we call the variables directly $\zeta$. The cancellation of the tadpole terms automatically guarantees the cancellation of the non-1PI diagrams, that otherwise should be included~(see Fig.~\ref{tadpole_2_point_cancellation}).

%%%%%%%%%%%%%%%%%%%%%%%%%%%%%%%%%%%%%%%%%%%%%%%%%%%%%%%%%%%%%%%%%%%%%%%%%%%%
%%%%%%%%%%%%%%%%%%%%%%%%%%%%%%%%%%%%%%%%%%%%%%%%%%%%%%%%%%%%%%%%%%%%%%%%%%%%%
%%%%%%%%%%%%%%%%%%%%%%%%%%%%%%%%%%%%%%%%%%%%%%%%%%%%%%%%%%%%%%%%%%%%%%%%%%%%
%%%%%%%%%%%%%

In unitary gauge, the resulting tadpole operator in $\delta g^{00}$ is of the form
\be
{\cal S}_{Tad,counter}= \int d^4x \sqrt{-g}\;\left[- \delta g^{00} 3 M_3^4(t)\langle(\delta g^{00}){}^2\rangle \right]\ ,
\ee
But since this has the same form as the induced tadpole operator that we have from $\left(\delta g^{00}\right)^3$, then the resulting quadratic (and higher order) terms that we obtain by expanding $\sqrt{-g}M^4_3(t)$ will also cancel.  This removes the contribution from the quartic  operators that would induce a time dependence.  

\begin{figure}[h]
\begin{center}
\includegraphics[width=17cm]{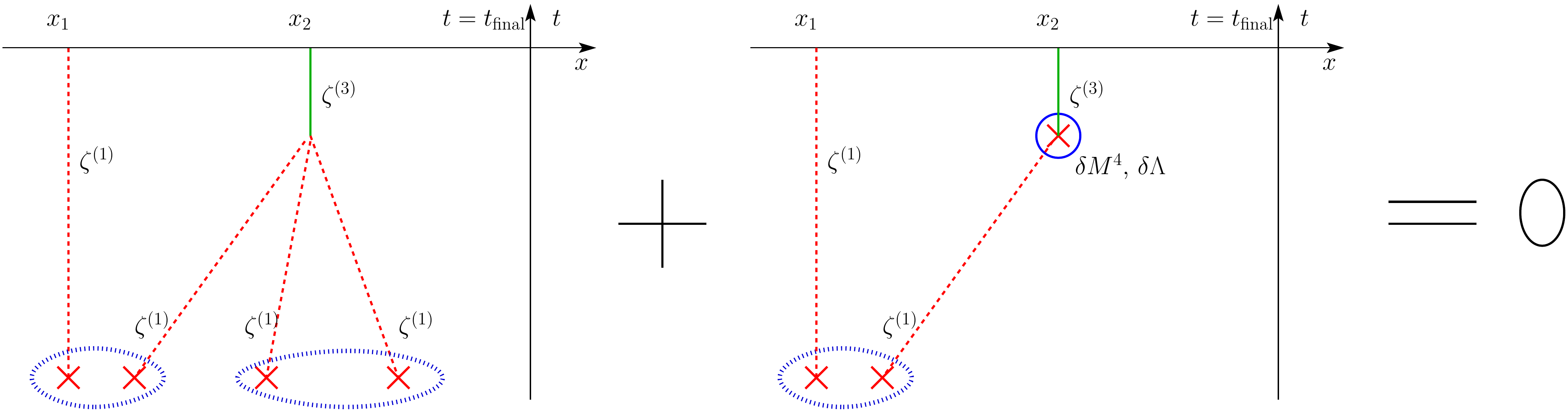}
\caption[]{ Cancellation of some quartic diagrams with the tree diagrams with an insertion of a counterterm-induced quadratic vertex.}
\label{quartic_cancellation}
\end{center}
\end{figure}

This can also be checked directly at the level of $\pi$. The dangerous term $\dot{M_3^4}\pi\dot\pi^3$ effectively gives a contribution that in the action can be represented as
\be
\delta {\cal S}_{4\rightarrow 2}=\int d^4x\sqrt{-g}\; \left[3 \dot M_3^4\pi\dot\pi\langle\dot\pi^2\rangle\right]\ ,
\ee
which is exactly cancelled by the tadpole term at second order
\be
{\cal S}_{Tad,counter}^{(2)}=\int d^4x\sqrt{-g}\; \left[-\dot\pi 3 M_3^4(t+\pi)\langle\dot\pi^2\rangle\right]\supset \int d^4x\sqrt{-g}\; \left[-3 \dot M_3^4(t)\pi\dot\pi\langle\dot\pi^2\rangle\right]\ .
\ee
This is represented in Fig.~\ref{quartic_cancellation}. Other quadratic terms induced by this tadpole operator are of the form $\dot\pi^2$ and $(\d_i\pi)^2$  and so do not induce time-dependent effects.
%%%%%%%%%%%%%%%%%%%%%%%%%%%%%%%%%%%%%%%%%%%%%%%%%%%%%%%%%%%%%%%%%%%%%%%%%%%%
%%%%%%%%%%%%%

This cancellation can be intuitively summarized by noticing that the $\zeta$ action at tree-level cannot have any mass term once expressed around the correct background. This is so because $\zeta$ constant must be a solution of the equations of motion when the mode is outside the horizon. The counterterms for tadpole diagrams ensure that  we are around the correct history, and so the quartic diagrams must cancel with the induced-quadratic diagrams from the tadpoles counterterms.

\section{Quartic diagrams: Verification for purely gravitational interactions}

Let us now move on and consider the most generic example for $H_4$ where we take generic coefficients and we do not neglect interactions mediated by gravity.  
Because of the complexity of this kind of interactions, the discussion becomes quite complicated even though all the essential points have already been highlighted using the Effective Field Theory of Inflation in the former section. We will therefore perform the study in several steps.

The first step will be to study the induced time dependence on the $\zeta$ zero mode, $\zeta_0$. As we discussed in eq.~(\ref{eq:diff1}) and~(\ref{eq:diff2}), the zero mode is not gauge fixed in the ordinary $\zeta$ gauge. We can fix the two functions in eq.~(\ref{eq:diff2}) in the following way: first we impose periodic boundary conditions. We imagine that the system is in a very large periodic box of comoving size $L$. In this way we forbid any dependence proportional to $x^i$. This fixes $\beta(t)$. Second, we can fix $C^i(t)$  by imposing that the zero mode component of $N^i$ vanishes: $N^i_{\vec k=0}(t)=0$.

\subsection{On the gauge choice for the zero mode}

Before proceeding, it is very interesting to notice the following. At finite $k$, $N^i$ is determined by being the solution of a constraint equation. At linear level, for example, the equation reads:
\be
\d_i N^i\sim \dot\zeta
\ee
which can be solved at finite $k$ to give
\be
N^i\sim \frac{k^i}{k^2}\dot\zeta \ .
\ee
In real space this term is often reported in a non-local fashion as $N^i\sim\frac{\d_i}{\d^2}\dot\zeta$. The zero momentum limit of that expression gives something that in real space reads as
\be
N^i(t)\sim \dot\zeta x^i\ .
\ee
By using our freedom in choosing the function $\beta$, we decided to set this term to zero. Therefore our solution for $N^i$ is {\it not} the $k\rightarrow 0$ limit of the solution for $N^i$ at finite $k$. We choose to work in a gauge where the limit is discontinuous. Of course any gauge choice should be as good as any other one. 

Working within the gauge where the limit $k\rightarrow 0$ of $N^i$ is continuous, that we can call `continuous gauge', raises several complications that we prefer to avoid. First of all, the continous gauge looks very unfamiliar when there is only a zero mode present. In this case the spacetime  is described by an FRW metric but the gauge choice makes us use  unusual coordinates where $g_{0i}\neq0$. But the situation becomes even more complicated. For example if in the continuous gauge we naively Taylor expand the action at linear level, we find that there is a tadpole term for the zero mode. The action starts linear, proportional to 
\bea
&&S=\frac{\mpl^2}{2}\int d^4x\,\sqrt{-g}\;\left[ R+\dot H \delta g^{00}+3 H^2+\dot H+\ldots\right] \quad \supset\\ \nonumber
&&\qquad\quad\sim\mpl^2\int d^4x\, a^3\, H \d_i N^i\sim\mpl^2 \int d^4x\; a^3\,\frac{\dot H}{H}\dot\zeta\ .
\eea
where $\ldots$ stands for terms that start explicitly quadratic in the fluctuations.
This is of course a wrong result, as the action for the fluctuations should start at quadratic order if we expand around a solution to the classical equations of motion, as we are doing. The reason for the mistake is that in this case the action has a boundary term that does not decouple in the limit in which we send the boundary to infinity. This is due to the behavior of $N^i\propto x^i$. Indeed the boundary term is the Gibbons-Hawking-York one:
\be
S_{GHY}=\mpl^2\int_{\d V^{(4)}}d^3\tilde x\;\sqrt{-h}\, K\ ,
\ee
where $h$ is the induced metric on the boundary described by coordinates $\tilde x$ and $K$ the trace of the extrinsic curvature. It is easy to check that this boundary term cancels the tadpole for the zero mode that we obtain from the bulk action.

The situation is instead much simpler in the `discontinuous gauge' where the limit $k\rightarrow 0$ of $N_i$ is discontinuous. In this case, for a fixed comoving box, the boundary terms become irrelevant as we send the boundary to infinity, and indeed the bulk action starts quadratic in the fluctuations. Furthermore, zero mode fluctuations appear to be directly in a standard FRW slicing. We will therefore work with this discontinuous gauge.

%{\bf ?!??!!?!?!?!?!?!?!?! is this enough?!?!?I am not putting the effect of the additional volume in the integral and so on.!?!?!?!?!?!!?!?!?!?!?!?!?!}

\subsection{Time-independence for the zero-mode}

We are now going to prove that the zero-mode is time-independent at one-loop. In order to do this, we need to expand the action to quadratic order in the zero-mode and independently up to quadratic order in the non-zero-modes. We count them as independent parameters. Since we expand only up to second order in each of the parameters, we need to solve the constraint solutions in the zero and in the short modes only at linear level in each of those. We work in Fourier space directly, and write
\bea
&&N=1+\delta N_k(t)+\delta N_0(t)\ , \\ \nonumber
&&N^i_{k}=\d_i \psi_k(t)\ .
\eea
We start from the action
\bea
&&S= \int d^3x\,dt\, \sqrt{h}\; \\
\nonumber
&&\left\{\frac{1}{2} \mpl^2
   \left(\frac{E_{ij} E^{ij}-E^i{}_i{}^2}{N}+N R \right)-\frac{ \mpl^2
   \dot H}{N}- N \mpl^2 \left(3
   H^2+\dot H\right)- N \delta \Lambda (t)-\frac{
   \delta M^4(t)}{N}\right\}\ ,
\eea
where the $\delta M^4$ and $\delta \Lambda$ terms represent the only two tadpole counterterms allowed by symmetries (all other possible choices are equivalent to those~\cite{Cheung:2007st}), and should be intended as objects that are of order $\zeta_k^2$.
The constraint equations read
\bea \nonumber
&&\frac{\mpl^2}{2}\left[R-\frac{1}{N^2}\left(E^i{}_jE^j{}_i-E^l{}_l\right)^2\right]+\frac{1}{N^2}\left(\mpl^2\dot H+   \delta M^4\right)-\left[\mpl^2\left(3 H^2+\dot H\right)+\delta\Lambda\right]=0\ ,\\
&&\hat\nabla_i\left[\frac{1}{N}\left(E^i{}_j-\delta^i{}_j E^l{}_l\right)\right]=0\ ,
\eea
and are solved by
\bea
&&\delta N_0(t)=\frac{3 H }{\dot H+3 H^2}\dot\zeta_0\ , \\ \nonumber
&&\delta N_k=\frac{(1+\delta N_0) }{H+\dot\zeta_0}\dot\zeta_k\ , \\ \nonumber
&&\psi_k=\frac{e^{-2 (\zeta_0+\rho (t))}}{k^2 \left(H+\dot\zeta_0\right)^2} \left(\dot H \dot\zeta_k e^{2
   (\zeta_0+\rho (t))}-(H+\dot\zeta_0)k^2 \zeta_k (t)
   ( 1+\delta N_0)^2\right)\ . \\ \nonumber
\eea
We plug back the above solutions into the action. At linear order the action is a total derivative, as it should be. At quadratic order, the zero-mode action reads
\be
S_{\zeta_0^2}=\int d^3x\, dt \;  e^{3 \rho (t)}\left(-\frac{3 \mpl^2 \dot H }{3
   H^2+\dot H}\right)\;\dot\zeta_0(t)^2\ ,
\ee
where we are writing $a(t)=e^{\rho(t)}$. It is interesting to notice that the quadratic action for the zero-mode is not the $k\rightarrow 0$ limit of the finite $k$ $\zeta$ action, the prefactor of $\dot\zeta_k^2$ being different. This is indeed
\be\label{eq:quadratci_short}
S_{\zeta_k^2}=\int d^3k\, dt \;  e^{3 \rho (t)}\left(-\frac{ \mpl^2 \dot H }{
   H^2}\right)\;\left(\dot\zeta_{\vec k}(t)\dot\zeta_{-\vec k}(t)- e^{-2\rho(t)}k^2\zeta_{-\vec k}\zeta_{\vec k}\right)\ .
\ee

\subsubsection{Tadpole Counterterms' Coefficients}

At this point we need to find the expressions for the tadpole counterterms $\delta\Lambda$ and $\delta M^4$ that ensure the cancellation of the tadpoles for $\zeta_0$. This is done by finding the cubic action at order $\zeta_0\zeta_k^2$, taking the expectation value on the short modes and canceling the resulting tadpole coefficients~\footnote{Notice that since we are working in the gauge $N^i_0=0$ and we choose a fixed comoving box in this gauge, there is no need to introduce boundary counterterms.}. Leaving out the simple algebra, the solution for the tadpole counterterms reads
\bea\nonumber\label{eq:tadpole-solutions}
\delta M^4&=&\frac{\mpl^2 e^{-2 \rho (t)} }{3 H^4}\left( H\left(-2
   \langle\d_i\zeta \d_i\dot\zeta\rangle \dot H+H (\langle\d_i\dot\zeta\d_i\dot\zeta\rangle+\langle\d_i\zeta\d_i\ddot \zeta\rangle)-H^2 \langle\d_i\zeta \d_i\dot\zeta\rangle+H^3 (-\langle\d_i\zeta \d_i\zeta\rangle)\right)\right. \\ 
   && \nonumber
  \left.\left.  -e^{2 \rho (t)}
   \left(6 H^2 \langle\dot\zeta\dot\zeta\rangle \dot H-3 \langle\dot\zeta\dot\zeta\rangle \dot H^2-9 H^3 \langle\zeta\dot \zeta\rangle
   \dot H+H \left(\langle\dot\zeta\dot\zeta\rangle \ddot H+2
   \langle\dot\zeta\ddot\zeta\rangle \dot H\right)+\right. \right. \right. \\
   && \left.\left.
   6 H^4
   (\langle\dot\zeta\dot\zeta\rangle+\langle\zeta\ddot \zeta\rangle)\right)\right)\ ,
   \\ \nonumber
   \delta \Lambda&=&\frac{\mpl^2 e^{-2 \rho (t)}}{3 H^4} \left( H \left(H (H
   (2 H \langle\d_i\zeta \d_i\zeta\rangle+5 \langle\d_i\zeta \d_i\dot\zeta\rangle)+\langle\d_i\dot\zeta\d_i\dot\zeta\rangle+\langle\d_i\zeta\d_i\ddot \zeta\rangle)-2 \langle\d_i\zeta\d_i\dot \zeta \rangle \dot H\right)\right.
   \\ \nonumber
  && \left.\left. -e^{2 \rho
   (t)} \left(3 H^2 \langle\dot\zeta\dot\zeta\rangle \dot H-3
   \langle\dot\zeta\dot\zeta\rangle \dot H^2+9 H^3 \langle\zeta \dot \zeta\rangle \dot H+H \left(\langle\dot\zeta\dot \zeta\rangle
   \ddot H+2 \langle\dot\zeta\ddot\zeta\rangle \dot H\right)\right.\right.\right. \\ \nonumber
   &&
\left.\left.  +6 H^4
   (\langle\dot \zeta\dot \zeta \rangle+\langle\zeta \ddot \zeta\rangle)+36 H^5 \langle\zeta\dot \zeta\rangle\right)\right)\ .
\eea
In these expressions, a term such as $\langle\d_i\zeta\d_i\zeta\rangle$ stands for $\langle\d_i\zeta(\vec x,t)\d_i \zeta(\vec x,t)\rangle$. A term like $\langle\zeta\dot\zeta\rangle$ stays for $\langle\zeta\dot\zeta+\dot\zeta\zeta\rangle/2$. No slow roll approximation has been performed nor it has ever been performed in this paper. There are three subtleties to stress here. The first is that the cubic action of order $\zeta_0\zeta_k^2$ is not the cubic action $\zeta_k^3$ with one of the momenta taken to zero. As before, the limit is discontinuous and the action is different. We do not report it here because it is very long and comes from trivial substitution of the solutions of the constraint equations into the action. Second, in taking expectation values $\langle\zeta^2\rangle$, one might worry about the contribution of the zero-mode, which has a different action than $\zeta_0\zeta_k^2$. This is irrelevant because the zero mode has measure zero when we perform the expectation value. The difference in the action is important for the tadpole terms and for the non 1-PI diagrams because the $\zeta_0$ propagator is the only one singled out by translation invariance. Finally, the third subtlety is about the expectation values involving two derivatives of $\zeta$: $\langle\zeta\ddot\zeta\rangle$. Here one can use the linear equation of motion for the short modes as derived from $(\ref{eq:quadratci_short})$ to relate it to expectation values of the form $\langle\zeta\d^2\zeta\rangle$ or $\langle\zeta\dot\zeta\rangle$.

\subsubsection{Cancellation between quartic diagrams and diff.-enhanced tadpole counterterms}

At this point we are able to address the time (in)dependence of the zero mode two-point function. In the former section we have discussed the contribution of the diagrams involving two cubic terms. We saw that upon the addition of some quartic diagrams, they induced no time dependence on $\zeta$. We have now to deal with the remaining quartic diagrams, that in this case come from the action of the form $\zeta_0^2\zeta_k^2$. 

The simplest way to evaluate the contribution of these diagrams to the $\zeta_0$ two-point function is to derive the quartic action and substitute directly the quadratic pieces in the short modes with their expectation value. For example
\be
\int d^3k\, dt\, e^{3\rho(t)}\; \zeta_0(t)^2\zeta_{\vec k}\zeta_{-\vec k}\quad\rightarrow\quad \int dt\, e^{3\rho(t)}\; \zeta_0(t)^2\langle\zeta^2\rangle\ ,
\ee
and then derive the resulting linear equation of motion for $\zeta_0$. In this way we can incorporate the effect of this quartic diagrams by simply studying the corrections to the quadratic action. The symmetries of the problem imply that the quadratic action will have a kinetic term  $\dot\zeta_0^2$ and a mass terms $\zeta_0^2$. There is also a term proportional to $\dot\zeta_0\zeta_0$ that can be reduced to a mass term upon integration by parts. Clearly a time dependence on $\langle\zeta_0^2\rangle$ can come only from a non vanishing mass term. These terms read
\bea
&&S^{(4)}_{\zeta_0^2,\;\zeta_0\dot\zeta_0}=\\ \nonumber
&&\int d t \left[\frac{\zeta_0^2}{2
   H^2} \left(
   \mpl^2  H e^{\rho (t)} (H \langle\d_i\zeta\d_i \zeta\rangle+2 \langle\d_i\zeta \d_i\dot\zeta\rangle)-9 e^{3 \rho (t)}
   \left(\mpl^2 \dot H \left(3 H^2  \langle\zeta
   \zeta\rangle+ \langle\dot\zeta\dot\zeta
   \rangle\right)\right.\right.\right.\\ \nonumber 
   &&
  \qquad\qquad \left.\left. +H^2 \left(3 \mpl^2 H \left(3H
   \langle\zeta \zeta\rangle+2 
   \langle\zeta \dot\zeta\rangle\right)+2 (\delta \Lambda
   +\delta M^4)\right)\right)\right)\\ \nonumber
&&\qquad +\frac{\zeta_0\dot\zeta_0}{ H^3 \left(\dot H+3
   H^2\right)} \left(
   \mpl^2 H e^{\rho (t)} \left(3 H^3 \langle\d_i\zeta
  \d_i \zeta\rangle-2 \langle\d_i\zeta \d_i\dot\zeta\rangle \dot H\right)\right.\\ \nonumber 
  &&\left.\qquad\qquad -3 e^{3 \rho
   (t)} \left(3 \mpl^2 H^2 \dot H \left(2 H
   \langle\zeta \dot \zeta \rangle+3 H^2  \langle\zeta
   \zeta \rangle-3 \langle\dot\zeta\dot\zeta\rangle\right)\right.\right. \\ \nonumber 
   &&\left.\left.\left.\qquad\qquad-2 \mpl^2  \langle\dot\zeta\dot\zeta\rangle \dot H^2+3 H^4 \left(9 \mpl^2 H^2 \langle\zeta \zeta\rangle+2 (\delta \Lambda-\delta M^4)\right)\right)\right)\right]\ .
  \eea
After we substitute in the counterterm solutions from (\ref{eq:tadpole-solutions}), and we integrate by parts the term $\zeta_0\dot\zeta_0$, the above expression simplifies to
\bea\label{S4remianing}
&&S^{(4)}_{\zeta_0^2}=\int dt \frac{ \mpl^2 e^{-\rho (t)} }{ H^2 \left(\dot H+3 H^2\right)^2}\ \zeta_0^2\\ \nonumber
&&\qquad\qquad\left(2 
   \langle\d_i\d_j\zeta \d_i\d_j\zeta\rangle \dot H \left(\dot H+3 H^2\right)+e^{2 \rho
   (t)} \left(2 \dot H \left(\dot H (H (H \langle\d_i\zeta \d_i\zeta \rangle+7
   \langle\d_i\zeta \d_i\dot\zeta\rangle)-\langle\d_i\dot\zeta\d_i\dot\zeta\rangle)\right.\right.\right.\\ \nonumber 
   &&\qquad\qquad\left.\left.\left.-3
   H^2 (H (2 H \langle\d_i\zeta \d_i\zeta\rangle-\langle\d_i\zeta \d_i\dot\zeta\rangle)+\langle\d_i\dot\zeta\d_i \dot\zeta\rangle)\right)+\ddot H \left(2
   \langle\d_i\zeta\d_i \dot \zeta\rangle\dot H-3 H^3 \langle\d_i\zeta \d_i\zeta\rangle\right)\right)\right)\ .
\eea
Clearly, a mass term seems to have survived after we have taken into account of the quadratic terms generated by the counterterm solutions. Unless these remaining terms are exactly those quartic terms of eq.~(\ref{quartic_spatial}), the terms associated with a rescaling of the spatial coordinates in cubic vertices, we would have a time-dependence for the $\zeta_0$ two point function. Luckily~\footnote{Or obviously, depending on the point of view.}, this is exactly what happens. It is indeed indicative that all the surviving terms have spatial derivatives acting on the $\zeta$'s inside the expectation values, suggesting that they are indeed associated to a rescaling of the spatial coordinates. Let us therefore discover what are those terms in (\ref{quartic_spatial}) by first finding the cubic Lagrangian of order~$\zeta_0\zeta_k^2$ and then taking the expectation value of the finite-$k$ modes. With the usual procedure, we obtain
\bea
&& S^{(3)}_{\zeta_0\zeta_k^2}=\int d^3x\;dt\;\left(-\frac{e^{\rho (t)}}{ H^3
   \left(\dot H+3 H^2\right)}\right)
 \\ \nonumber
&&\qquad \left(H^3 \dot H \left(\mpl^2 \left(\zeta_0 \left(- \left(\langle\d_i\zeta\d_i \zeta\rangle-9 e^{2
   \rho (t)}\langle\zeta \zeta\rangle\dot H\right)+12 e^{2 \rho (t)} \dot H+9 \langle\dot\zeta \dot\zeta\rangle e^{2 \rho (t)}\right)+6 \langle\zeta \dot\zeta\rangle e^{2
   \rho (t)} \dot\zeta_0\right)\right.\right. \\ \nonumber
   &&\qquad\left.\left.+6 \delta \Lambda (t) \zeta_0 e^{2 \rho (t)}+6 \delta M^4 \zeta_0 e^{2
   \rho (t)}\right)-3 H^4 \left(\dot\zeta_0 \left(\mpl^2
   \left( \left(\langle\d_i\zeta \d_i\zeta\rangle-3 e^{2 \rho (t)}
 \langle\zeta \zeta\rangle  \dot H\right)-4 e^{2 \rho (t)} \dot H\right)\right.\right.\right.\\ \nonumber
   &&\left.\left.\left. \qquad-2 \delta \Lambda e^{2 \rho
   (t)}+2 \delta M^4 e^{2 \rho (t)}\right)+2 \mpl^2
   \zeta_0  \left(\langle\d_i\zeta \d_i\dot\zeta\rangle-3
   e^{2 \rho (t)} \langle\zeta \dot\zeta\rangle\dot H \right)\right)\right.\\ \nonumber 
   &&\qquad\left.+3 H^5 \zeta_0
   \left(\mpl^2 \left(-\left(
   \left(\langle\d_i\zeta\d_i \zeta\rangle-18 e^{2 \rho (t)}\langle\zeta \zeta\rangle\dot H\right)-24 e^{2 \rho (t)}
   \dot H\right)\right)+6 \delta \Lambda e^{2 \rho (t)}+6 \delta M^4 e^{2 \rho (t)}\right)\right. \\ \nonumber 
   &&\qquad \left.-\mpl^2 H^2 \dot H \left(2
    \zeta_0 \langle\d_i\zeta \d_i\dot \zeta\rangle+9
   \langle\dot\zeta\dot\zeta\rangle e^{2 \rho (t)} \dot\zeta_0\right)+\mpl^2 H  \dot H \left(3 \zeta_0
   \langle\dot\zeta\dot\zeta\rangle e^{2 \rho (t)} \dot H+2 k^2
   \langle\zeta \dot \zeta\rangle \dot\zeta_0\right)\right.\\ \nonumber 
   && \left.-2 \mpl^2
   \langle\dot\zeta\dot\zeta\rangle e^{2 \rho (t)} \dot H^2 \dot\zeta_0+9 \mpl^2 H^6 e^{2 \rho (t)} \left(3 \langle\zeta \zeta\rangle \dot\zeta_0+6 \zeta_0 \langle\zeta\dot \zeta\rangle\right)+81 \mpl^2 H^7 \zeta_0 
   \langle\zeta \zeta\rangle e^{2 \rho (t)}\right)\ .
   \eea
According to the results of sec.~\ref{sec:cubic}, loops formed with cubic operators that contain spatial derivatives would induce  time dependence unless we combine them with quartic loops constructed with the operators derived from formula~(\ref{quartic_spatial}).  Applying it to the cubic action above, we obtain
\bea
&&S_{Quartic,\d_i}=\int d^3x\;dt\; \left(-\frac{ \mpl^2  e^{\rho
   (t)} }{ H^2 \left(\dot H+3 H^2\right)}\zeta_0\right) \\ \nonumber
&&\qquad\qquad \left(-4 \langle\d_i\zeta \d_i\dot\zeta \rangle \dot H \dot\zeta_0+2
   H \zeta_0 \langle\d_i\zeta\d_i\dot \zeta\rangle \dot H+H^2
   \zeta_0 \langle\d_i\zeta \d_i\zeta\rangle \dot H+6 H^3
   \langle\d_i\zeta\d_i \zeta \rangle \dot \zeta_0\right.\\ \nonumber 
   &&\qquad\qquad\qquad\left.+6 H^3 \zeta_0 \langle\d_i\zeta\d_i \dot\zeta\rangle+3 H^4 \zeta_0
   \langle\d_i\zeta\d_i \zeta\rangle\right) \ .
\eea
Upon integration by parts, and after using the equation of motions in terms of the form $\langle\ddot\zeta\zeta\rangle$, it is easy to see that these terms are exactly the ones left out in (\ref{S4remianing}). Notice that we do not even need to compute explicitly the value of $\langle\d\zeta\d\zeta\rangle$: it cancels with the corresponding terms. This shows that one can combine the terms in (\ref{S4remianing}) with the diagrams built with cubic interactions to see that all those diagrams do not give a time dependence to $\zeta_0$. The remaining quartic diagrams cancel with the quadratic terms induced by the tadpole terms. 

This concludes all the diagrams that appear at one loop. We see that both the 1-PI and non 1-PI diagrams are important to cancel each other so that, even though naively many diagrams are dangerous and can potentially give a time dependence to the $\zeta_0$ correlation function, the time-dependence cancels in the sum, and we conclude that the $\zeta_0$ two-point function is time independent.

\subsection{Time-independence for the non-zero-modes}

We are now ready to begin the study of the case in which the external momentum is finite. This task is very challenging~\footnote{At least for our standards.}, as the interactions are even more complicated than for the case of the zero mode. Luckily we will be able to do it by employing a trick. As we discussed, the time-dependence we are interested in ruling out is the one that appears when the wavelength of the mode is much longer than the horizon, and the loop effect is due to short wavelength modes running in the loop (modes longer than our mode clearly  cannot induce a time dependence). For this reason, we can simplify the action by taking the leading term in the smallness of the derivatives of the external mode. 

In $\zeta$-gauge, this simplification is not trivial at all. After substituting the solutions to the constraint equations, $N^i$ becomes larger and larger as we move to finite but smaller and smaller $k$'s. This is due to the fact that at finite $k$, $N^i$ has the non-local-looking expression $N^i\sim k^i\dot\zeta/k^2\ $~\footnote{We stress that since we are trying to investigate if $\zeta_k$ becomes time-dependent, we cannot assume that $\dot\zeta\sim k^2\zeta/a^2$ out of the horizon, as it happens in the free theory. Indeed time derivatives do not count as a suppression when the mode is part of a commutator in a Green's function.}. Armed with the experience of the zero-mode, we realize that it would probably be much better if we could find a gauge where $N^i$ does not have this bad behavior at low momenta. Since at finite $k$ all gauge freedoms are completely fixed by the $\zeta$-gauge conditions, this is globally impossible. However, we can do this locally. Indeed, we can find a frame valid in a region of space very small compared to the wavelength of the mode, where the universe looks like an anisotropic flat universe. 
%We will define short fluctuations directly in this frame and perform a loop over them. 
Corrections to the results obtained in this frame will be down by powers of $k/(aH)$ and so will lead to a contribution that is convergent with time. Since we are dealing with a time-dependent finite-$k$ Fourier mode, the local frame is not a local FRW universe as it was for the zero mode, but it is an anisotropic universe. For simplicity, we can choose to work directly with a single Fourier mode
\be
\zeta_k(\vec x, t)={\rm Re}\left[\tilde \zeta_0(t) \;e^{i \vec k\cdot \vec x}\right]\ ,\qquad {\rm Re}\left[\tilde \zeta_0\right]=\zeta_0\ .
\ee
Using rotational invariance, we can take the momentum $\vec k$ to be along the $\hat z$ direction without loss of generality.  The resulting spatial metric in the ADM parametrization is given by the following:
\bea\label{eq:anisotropic_metric}
&& \hat h_{11}=\hat h_{22}=e^{2\rho(t)+2\zeta_0(t)+2\lambda_0(t)} e^{2\zeta(\vec x,t)}\ , \\ \nonumber
&& \hat h_{33}=e^{2\rho(t)+2\zeta_0(t)-4\lambda_0(t)} e^{2\zeta(\vec x,t)}\ , \\ \nonumber
&& N_i=\d_i\psi(\vec x,t)+\tilde N_i(\vec x,t) \ , \qquad \d^i\tilde N_i(\vec x,t)=0 \\ \nonumber
&& N=1+\delta N_0(t)+\delta N(\vec x,t)\ .
\eea
Here the fields with the argument $\vec x$ represent short wavelength fields that will  be integrated over in the loops. We see that there is no $N_{i,0}(t)$ component. This is so because we can make $N^i_0$ and $\d_i N^j_0$ vanish. The field $\lambda_0$ is the (traceless) anisotropic component of the metric, related to $\zeta_0$ by
\be
\lambda_0(t)=-\frac{1}{3} \int^tdt'\;\frac{\dot H}{ H^2}\dot \zeta_0\ ,
\ee
up to an irrelevant constant that can be set to zero using a constant rescaling of the spatial coordinates. The details of this change of coordinates are given in App.~\ref{app:local_frame}.

Apart from the terms proportional to $\lambda_0$, the treatment is very parallel to the one of the former subsection. First we find the solution to the tadpole counterterms $\delta M^4$ and $\delta \Lambda$. As expected, there is no tadpole for the terms in $\lambda_0$ because of rotational invariance: the free vacuum expectation value of product of fields must be rotational invariant and cannot source any anisotropy. This is indeed the case, and the solutions for $\delta M^4$ and $\delta \Lambda$ are exactly the same as before~eq.~(\ref{eq:tadpole-solutions})~\footnote{There is only one subtlety here that distinguishes this case from the former one. In the former section we were studying the effect of loops on the zero mode, and therefore loop integrals whose range is over momenta that are shorter than the external one, were basically running over all momenta. Here instead, since we are Taylor expanding in derivatives of the long external mode, loops should formally include only modes that are shorter than the external one. This is hardly a problem however because in order to prove that there is no induced time-dependence, we are interested in the case where the external mode $k$ is outside of the horizon. The contribution from modes longer than the horizon is equivalent to the contribution of modes that are all out of the horizon. At this point, a nice property of the $\zeta$ action tells that there are no vertices without at least a derivative acting on one $\zeta$ fluctuation~\cite{Maldacena:2002vr}. This guarantees that when all the modes are outside of the horizon each vertex is suppress by powers of $k/(aH)$. So those contributions would give rise to a time-convergent contribution and can be safely ignored.}. 

At this point we proceed to find the action for the short modes in this background. We start with the solution to the constraint equations, that read:
\bea
&&\delta N_0(t)=\frac{3 H }{3 H^2+\dot H} \dot\zeta_0\ , \\ \nonumber
&& \delta N_k=\frac{\dot \zeta_k}{H^2}  \left(H \delta N_0+H-\dot\zeta_0\right)+\frac{k_{\rm ani}^2}{2 k^2 H^2}\left(\dot\zeta_k-3 H\zeta_k\right)\dot\lambda_0\ ,\\ \nonumber
&&\psi_k=\frac{e^{2 \rho (t)}}{2 k^4 H^3} \left[2\dot H \dot\zeta_k \left(H \left(2
   k_{\rm ani}^2 \lambda_0+2 k^2 \zeta_0+k^2\right)-2 k^2 \dot\zeta_0\right)+\right.  \\ \nonumber
   &&\qquad\dot\lambda_0\left(-3 H\zeta_k\left(3H^2+\dot H\right)+\left(3 H^2+2\dot H\right)\dot\zeta_k\right)+\\ 
   \nonumber &&\qquad\left.
    k^2 H \zeta_k
   \left(-k_{\rm ani}^2 \dot\lambda_0-2 k^2
   \left(2 H \delta N_0+H-\dot\zeta_0\right)\right)\right]\ , \\ \nonumber
 && \tilde N_i=\frac{ k_i e^{2 \rho (t)}}{k^4 H }2 \left( k^2- k_{\rm ani}^2\right)
   \dot\lambda_0 \left(3 H \zeta_k -\dot\zeta_k\right)\ , \quad i=1,2\ , \\ \nonumber
   && \tilde N_3=\frac{ k_3 e^{2 \rho (t)}}{k^4 H }2 \left(2 k^2- k_{\rm ani}^2\right)
   \dot\lambda_0 \left(3 H \zeta_k -\dot\zeta_k\right)\ , 
\eea
where $k^2=k_x^2+k_y^2+k_z^2$ and $k_{\rm ani}^2=k_x^2+k_y^2-2k_z^2$. $k_{\rm ani}$ has the nice property that $\int d^2\hat k \;k_{\rm ani}^2=0$. After substitution of the above solutions in the action, we obtain the quartic action at order $\zeta_0^2\zeta_k^2$. As before we evaluate the expectation value on the $\zeta_k$-modes and isolate  the terms in $\zeta_0$ (and $\lambda_0$) that could lead to a time-dependence for $\zeta_0$. Clearly, we need to keep track only of the terms that contain at least one $\lambda_0$, the terms quadratic in $\zeta_0$ will cancel exactly as in the former section. Furthermore, because of rotational invariance, terms proportional to $\lambda_0\zeta_0$ are absent. We are left with
\bea
&&S^{(4)}_{\lambda_0^2}= \int d^3x\,dt\; e^{\rho (t)}\lambda_0^2\frac{4 \mpl^2 }{H} \left(H \langle \d_i\zeta\d_i\zeta\rangle+2 \langle \d_i\zeta\d_i\dot\zeta\rangle\right)\ , \\ \nonumber
  &&S^{(4)}_{\lambda_0\dot\lambda_0}= \int d^3x\,dt\; \left(-2e^{\rho (t)}\frac{\mpl^2}{ H^3} \right)
   \left(-3 H\left(\langle\frac{\d_{\rm ani}^2}{\d^2}\zeta\frac{\d_{\rm ani}^2}{\d^2}\dot\zeta\rangle-2 \langle\zeta\dot\zeta\rangle\right) e^{2 \rho (t)} \dot H
   -6 H^2 \langle\d_i\zeta\d_i\zeta\rangle\right.\\ \nonumber 
   &&\quad + \left.\left( \langle\frac{\d_{\rm ani}^2}{\d^2}\dot\zeta\frac{\d_{\rm ani}^2}{\d^2}\dot\zeta\rangle-2 \langle\dot\zeta\dot\zeta\rangle\right) e^{2 \rho (t)} \dot H
  +2  H \langle\d_i\zeta\d_i\dot\zeta\rangle\right)\ .
\eea
Here $\d^2=\d_x^2+\d_y^2+\d_z^2$ while $\d^2_{\rm ani}=\d_x^2+\d_y^2-2 \d_z^2$. The second expression above can be further simplified by noticing that by rotational invariance 
\be
\langle\frac{\d_{\rm ani}^2}{\d^2}\dot\zeta\frac{\d_{\rm ani}^2}{\d^2}\dot\zeta\rangle=\frac{4}{5}\langle\dot\zeta\dot\zeta\rangle\ ,
\ee
and similar for similar terms. After integrating by parts the term in $\lambda_0\dot\lambda_0$ and summing with the term in $\lambda_0^2$, we obtain the final expression
\bea
&&S^{(4)}_{\lambda_0^2}= \int d^3x\,dt\; e^{-\rho (t)}\;\lambda_0^2\frac{2\mpl^2 }{5
   H^4 \dot H}\\ \nonumber
   &&\qquad  \left(-20
    H^3 e^{2 \rho (t)} \dot H\langle \d_i\zeta \d_i\dot \zeta\rangle-5
    H^4 \langle\d_i\zeta\d_i\zeta\rangle e^{2 \rho (t)} \dot H-3 e^{4 \rho (t)}
   \dot H^3 \langle \dot \zeta\dot\zeta\rangle\right. \\ \nonumber
   &&\left.\qquad +H^2 \left(-5  e^{2 \rho
   (t)} \ddot H \langle \d_i\zeta\d_i\dot\zeta\rangle+e^{2 \rho (t)} \dot H  \left(18 e^{2
   \rho (t)} \dot H \langle\dot\zeta \dot\zeta\rangle+5 \langle\d_i\dot\zeta \d_i\dot\zeta\rangle\right)\right.\right.\\ \nonumber
   &&\left.\left.\qquad- \dot H \left(5  \langle\d_j\d_i\zeta\d_j\d_i\zeta\rangle-6 e^{2 \rho (t)}
   \dot H \langle\d_i\zeta\d_i\zeta\rangle\right)\right)+3 H e^{2 \rho (t)} \dot H \left(e^{2 \rho
   (t)} \ddot H \langle \dot\zeta\dot\zeta\rangle+2  \langle\d_i\zeta\d_i \dot\zeta\rangle \dot H\right)\right)\ .
\eea
As in the former section, if these terms were not to be exactly the ones in Quartic$_{\d_i}$ then we will have a time dependence for the $\zeta$ correlation function. To check for this, we move to the cubic action. Again, we need simply to investigate terms proportional to $\lambda_0\zeta_k^2$. We have
\bea
&& S^{(3)}_{\lambda_0\zeta_k^2}=\int d^3x\,dt\;2\frac{ \mpl^2 e^{\rho (t)}}{
   H^3} \\ \nonumber
&&\qquad  \left(e^{2 \rho (t)} \dot H
    \dot\lambda_0 \left(3 H\frac{\d_{{\rm ani}}^2}{\d^2}\zeta\dot\zeta-\frac{\d_{{\rm ani}}^2}{\d^2}\dot\zeta\dot\zeta\right)- H \left(\dot\lambda_0
   \left(\d_{{\rm ani}}^2\zeta\dot\zeta-3 H \d_{{\rm ani}}^2\zeta\zeta\right)\right.\right.\\ \nonumber
   &&\qquad\left.\left.-2 H \lambda_0
   \left(H \d_{{\rm ani}}^2\zeta\zeta+2 \d_{{\rm ani}}^2\zeta\dot\zeta\right)\right)\right)\ .
\eea
The identification of the quartic vertices starting from the cubic vertices is slightly more complicated due to the anisotropy. In practice, everytime in the cubic Lagrangian there are two derivatives that are contracted, they should be thought of as originating from being contracted with the spatial metric  $\hat h_{ij}$, and we take the resulting relevant quartic operator. Let us give a few examples:
\bea
{\cal L}_3 \quad\supset \quad \zeta_0(\d_i\zeta)^2 \qquad&\rightarrow& \qquad  {\cal L}_4 \quad\supset\quad -\zeta_0^2(\d_i\zeta)^2-2\lambda_0\zeta_0 (\d_{\rm ani}\zeta)^2\ , \\ \nonumber
{\cal L}_3 \quad\supset \quad \dot\zeta_0(\d_i\zeta)^2 \qquad&\rightarrow& \qquad {\cal L}_4 \quad \supset\quad-  2\zeta_0\dot\zeta_0(\d_i\zeta)^2-  2\lambda_0\dot\zeta_0(\d_{\rm ani}\zeta)^2\ ,\\ \nonumber
{\cal L}_3 \quad\supset \quad \lambda_0(\d_i\zeta)^2 \qquad&\rightarrow& \qquad {\cal L}_4 \quad \supset\quad-  2\zeta_0\lambda_0(\d_i\zeta)^2-  \lambda_0^2(\d_{\rm ani}\zeta)^2\ ,\\ \nonumber
{\cal L}_3 \quad\supset \quad \dot\lambda_0(\d_i\zeta)^2 \qquad&\rightarrow& \qquad {\cal L}_4 \quad \supset\quad-  2\zeta_0\dot\lambda_0(\d_i\zeta)^2-  2\lambda_0 \dot\lambda_0(\d_{\rm ani}\zeta)^2 ,\\ \nonumber
{\cal L}_3 \quad\supset \quad \zeta_0(\d_{\rm ani}\zeta)^2 \qquad&\rightarrow& \qquad  {\cal L}_4 \quad\supset\quad (-\zeta_0+2\lambda_0)\zeta_0(\d_{\rm ani}\zeta)^2-4\lambda_0 \zeta_0(\d_{i}\zeta)^2\ , \\ \nonumber
{\cal L}_3 \quad\supset \quad \dot\zeta_0(\d_{\rm ani}\zeta)^2 \qquad&\rightarrow& \qquad  {\cal L}_4 \quad\supset\quad (-2\zeta_0+2\lambda_0)\dot\zeta_0(\d_{\rm ani}\zeta)^2-4\lambda_0 \dot\zeta_0(\d_{i}\zeta)^2\ , \\ \nonumber
{\cal L}_3 \quad\supset \quad \lambda_0(\d_{\rm ani}\zeta)^2 \qquad&\rightarrow& \qquad  {\cal L}_4 \quad\supset\quad (-2\zeta_0+\lambda_0)\lambda_0(\d_{\rm ani}\zeta)^2-2\lambda_0^2(\d_{i}\zeta)^2\ , \\ \nonumber
{\cal L}_3 \quad\supset \quad \dot\lambda_0(\d_{\rm ani}\zeta)^2 \qquad&\rightarrow& \qquad  {\cal L}_4 \quad\supset\quad (-2\zeta_0+2\lambda_0)\dot\lambda_0(\d_{\rm ani}\zeta)^2-4\lambda_0 \dot\lambda_0(\d_{i}\zeta)^2\ ,
\eea
where $\vec\d_{{\rm ani}}=(\d_x,\d_y,i\sqrt{2}\d_z)$.
Upon implementing the promotion of the spatial derivative to include the $\zeta_0$ and $\lambda_0$ factors, we have
\bea
&&S_{Quartic,\d_i}=\int d^3x\,dt\; e^{\rho (t)}\,  \lambda_0\, \frac{4 \mpl^2 }{5
   H(t)^3}\\ \nonumber
&&\qquad  \left(5
    H \left(\dot\lambda_0 \left(3 H
   \langle \d_i\zeta\d_i\zeta\rangle-\langle \d_i\zeta\d_i\dot\zeta\rangle\right)+H \lambda_0 \left(H \langle\d_i \zeta\d_i\zeta\rangle
   +2 \langle\d_i \zeta\d_i\dot\zeta\rangle\right)\right)-\right. \\ \nonumber 
   &&\left.\qquad 3 e^{2 \rho (t)} \dot H 
   \dot\lambda_0 \left(3 H \langle\zeta\dot\zeta\rangle-\langle\dot\zeta\dot\zeta\rangle\right)\right)= \\ \nonumber
  && \quad= \int d^3x\,dt\; e^{-\rho (t)}\;\lambda_0^2\frac{\mpl^2 }{5
   H^4 \dot H}\\ \nonumber
   &&\qquad  \left(-20
    H^3 e^{2 \rho (t)} \dot H\langle \d_i\zeta \d_i\dot \zeta\rangle-5
    H^4 \langle\d_i\zeta\d_i\zeta\rangle e^{2 \rho (t)} \dot H-3 e^{4 \rho (t)}
   \dot H^3 \langle \dot \zeta\dot\zeta\rangle\right. \\ \nonumber
   &&\left.\qquad +H^2 \left(-5  e^{2 \rho
   (t)} \ddot H \langle \d_i\zeta\d_i\dot\zeta\rangle+e^{2 \rho (t)} \dot H \left(18 e^{2
   \rho (t)} \dot H \langle\dot\zeta \dot\zeta\rangle+5  \langle\d_i\dot\zeta\d_i \dot\zeta\rangle\right)\right.\right.\\ \nonumber
   &&\left.\left.\qquad- \dot H \left(5  \langle\d_j\d_i\zeta\d_j\d_i\zeta\rangle-6 e^{2 \rho (t)}
   \dot H \langle\d_i\zeta\d_i\zeta\rangle\right)\right)+3 H e^{2 \rho (t)} \dot H \left(e^{2 \rho
   (t)} \ddot H \langle \dot\zeta\dot\zeta\rangle+2  \langle\d_i\zeta\d_i \dot\zeta\rangle \dot H\right)\right)\ .
   \eea
 where in the first passage we have used that by rotational invariance terms involving $\d_{\rm ani}^2$ are zero and those involving $\d_{\rm ani}^4$ are equal to the same expression with $(\d_{\rm ani}^2)^2\rightarrow 4(\d^2)^2/5$, and in the second passage we have performed an integration by parts. We see that this $Quartic_{\d_i}$ term is exactly the one being left out from the loops with the quartic diagrams, and so its time-dependent contribution will cancel the one coming from the $CIS_{1PI}+CIM+Quartic_{\d_t}$ diagrams. This completes the exploration of  all the diagrams entering at one-loop, proving that the $\zeta_k$ correlator does not have a time-dependence even at finite momentum $k$.

\vspace{0.6cm}
{\bf A note on tensor modes:} Since in this section we have dealt with gravitational interactions, it is logical to wonder on the contribution of the tensor modes. Indeed, for standard slow roll inflation, at one-loop the contribution from tensor modes is parametrically the same as the one from the $\zeta$ short modes. One might wonder why we could neglect them, or alternatively why time-dependent effects from  loops of $\zeta$ modes cancel independently of the ones from loops of  tensor modes.  It is easy to realize that the contribution from tensor modes must cancel independently. Let us analyze the various diagrams. It  is pretty clear that the diagrams built with cubic vertices will cancel independently in the same way as they independently did for the $\zeta$ modes. This cancellation  in fact relies on the consistency condition, that holds for tensor modes as well as for $\zeta$ modes. A bit less obvious is to understand why the graviton and $\zeta$ contribution from quartic and tadpole terms cancel independently. The fact that the contribution of tensor modes and scalar modes is parametrically the same is an accident of standard slow roll inflation. It is possible to engineer inflationary models where the contribution is parametrically different. If for example we add to the Effective Field Theory of Inflation an operator of the form $(\delta g^{00}){}^2$, we change the speed of sound of the $\zeta$ fluctuations, without changing the ones of the tensor modes. Since the tadpoles and the quartic loops are evaluated on the linear solutions, this shows that those loops are parametrically different, and they have to cancel independently. We have explicitly verified that this is the case for the effect on the $\zeta$ zero-mode.

\section{Conclusions} 

Understanding the behavior of the theory of inflationary fluctuations at one-loop order, with particular attention to the possible infrared factors, is a very important task. We have stressed how this is important for the predictivity of inflation as well as for slow roll eternal inflation and its universal volume bound. In general, it is also important to understand how the theory we believe to be the strongest contender for describing the first instants in the history of our universe behaves at quantum level.  

In this paper we have proven that the $\zeta_k$ correlation function does not receive corrections that grow with time $\sim Ht$ after the mode has crossed the horizon. This result is achieved by proving that there is a cancellation among the various diagrams that would naively induce a time-dependence, if taken alone. While this cancellation happens in an intricate way, its physical origin can be stated in a very simple form. First, since there is a vacuum contribution to the stress tensor due to the fluctuations, it is important to define the $\zeta$ fluctuations around the correct one-loop spacetime background. This can be achieved either by automatically including $non-1PI$ diagrams in the calculation, or, as we do here, by inserting  diff. invariant counterterms that cancel the tadpole correction. Because of diff. invariance, these tadpole counterterms contain terms quadratic in the fluctuations that modify the $\zeta$ propagator and account for a cancellation of the time-dependence induced by many of the diagrams built from quartic vertex. Some of these quartic diagrams indeed look very much like coming from a renormalization of the background, as they involve vacuum expectation values of quadratic operators on the unperturbed background. It is not so surprising that they cancel with the tadpole counterterms.

The remaining quartic vertices, that we have called $Quartic_{\d_t}$ and $Quartic_{\d_i}$, induce a time dependence that cancels with the one from the cubic diagrams that we call $CIS_{1PI}+CIM$. The sum of all these diagrams describes how the vacuum expectation value of the short-wavelength modes is affected by the presence of a long-wavelength mode, and  how the perturbation in this expectation value in turn backreacts on the long-wavelength mode. Because of the attractor nature of the inflationary solution, a long wavelength $\zeta$ fluctuation is equivalent to a trivial rescaling of the coordinates in the unperturbed background. So the vacuum expectation value of the short-wavelength modes should not be affected at all by the presence of a long wavelength mode making this effect disappear.

Since the $\zeta$ fluctuations are not derivatively coupled, a feature shared also by the graviton, showing this is not easy. In order to do it we wrote the sum of these diagrams as the three-point function between two short-wavelength modes and one long-wavelength mode, integrated over the short-wavelength Fourier components. In this way, after adding the terms from $Quartic_{\d_t}$ and $Quartic_{\d_i}$, we could use the consistency condition to show that the presence of a long-wavelength $\zeta$ does not change the expectation value of short modes in a way that correlates with the long mode and therefore that these diagrams do not give any time dependence. 

By accounting for all the  diagrams at one loop order we proved that $\zeta$ is a constant at this order.

There are many possible generalizations to our results. In the introduction we gave arguments that could be easily generalized to arbitrary loops. Furthermore it would be nice to include in the treatment gravitons both inside the loops as well as in the external legs. All of this seems doable. The physical principles responsible for the cancellations we found should hold unchanged also for these more general cases.

\subsubsection*{Acknowledgments}

We thank Nima Arkani-Hamed, Sergei~Dubovsky, David Gross, Steve Giddings, Richard Holman, Shamit Kachru, Matt~Kleban, Juan Maldacena, Steve Shenker, Eva Silverstein, Lenny Susskind, Giovanni Villadoro and Steven Weinberg for interesting conversations. G.L.P. was supported by the Department of State through a Fulbright Science
and Technology Fellowship and through the US NSF under Grant No. PHY-0756966. L.S.~is supported by the National Science Foundation under PHY-1068380.
M.Z. is supported by the National Science Foundation under PHY-
0855425 and AST-0907969 and by the David and Lucile 
Packard Foundation and the John D. and Catherine~T.~MacArthur~Foundation. L.S. and M.Z. acknowledge the CERN Theory Institute `Quantum Gravity from UV to IR' where part of this work was performed. L.S. acknowledges hospitality of the Institute for Advanced Study where part of this work was performed. 

\begin{appendix}
\section*{Appendix}
\section{Consistency Condition inside the Horizon\label{app:consistency-inside}}

In this Appendix we discuss the three-point function in the squeezed limit in which one of the modes is much longer than the other two. While so far the literature has always concentrated in the limit in which the two short modes are outside of the horizon, as this is the relevant limit for observed modes in tree-level correlation functions, at loop level we are also interested in the case in which the two short modes are inside the horizon. We will verify that the consistency condition also holds in this regime. We will do this at leading order in slow roll parameters. 

For the case in which the short modes are still inside the horizon, the proof at leading order in slow roll parameters is very easy. In fact, contrary to what happens when we are interested in computing the correlation function of modes at a time when they are outside the horizon, in this case the leading interaction is of zero$^{th}$ order in the slow roll parameters. Indeed, it is not true that the $\zeta$ cubic action starts at first order in slow roll parameters (relative to the quadratic action). This is so only up to terms that can be removed by a field redefinition and that can therefore be evaluated at the final time. For modes that are outside of the horizon at the time of evaluation, these vanish. For modes that are not yet outside of the horizon, they do not, and they therefore represent the leading contribution in the slow roll expansion.

Following~\cite{Maldacena:2002vr}, the term we are discussing comes from the field redefinition:
\be
\zeta=\zeta_n+\frac{\zeta\dot\zeta}{H}+\ldots\ ,
\ee
where $\ldots$ represent terms suppressed by slow roll parameters. The variable $\zeta_n$ has a cubic action that is suppressed by slow roll parameters, and so negligible. At this point computing the three-point function is very straightforward. In the limit in which the long mode $k_3$ is much longer than the horizon $k_3/a(\eta)\ll H$ and $k_3\ll k_2\simeq k_1$, we have
\bea
&&\langle\zeta_{k_1}(\eta)\zeta_{k_2}(\eta)\zeta_{k_3}(\eta)\rangle\simeq (2\pi)^3\delta^{(3)}(\vec k_1+\vec k_2+\vec k_3) \frac{1}{H}\langle\dot\zeta_{k_1}\zeta_{k_1}+\zeta_{k_1}\dot\zeta_{k_1}\rangle'\;\langle\zeta_{k_3}^2\rangle'\\ \nonumber &&
\qquad=(2\pi)^3\delta^{(3)}(\vec k_1+\vec k_2+\vec k_3)\frac{1}{H} \d_t\langle\zeta_{k_1}^2\rangle'\;\langle\zeta_{k_3}^2\rangle'\ , \qquad\quad\ k_1\ll k_3\ ,
\eea
where the $\langle\rangle'$ symbol stays for the fact that we have removed the delta function from the expectation value. Using the wavefunction of the modes at leading order in slow roll parameters
\be\label{eq:zetacl}
\zeta^{cl}_k(\eta)=\frac{H}{2\sqrt{ \epsilon}\mpl}\frac{1}{k^{3/2}}\left(1-i k\eta\right)e^{i k \eta}\ ,
\ee
where $\epsilon$ is the slow roll parameter $\epsilon=-\dot H/H^2$, we obtain
\be\label{eq:consistency_inside}
\langle\zeta_{k_1}(\eta)\zeta_{k_2}(\eta)\zeta_{k_3}(\eta)\rangle\simeq-\frac{ H^4}{8  \mpl^4 \epsilon^2}\cdot\frac{\eta ^2}{k_1 k_3^3}\ .
\ee

In order to satisfy the consistency condition, the above result should be equal to
\be
\langle\zeta_{k_1}(\eta)\zeta_{k_2}(\eta)\zeta_{k_3}(\eta)\rangle\simeq- (2\pi)^3\delta^{(3)}(\vec k_1+\vec k_2+\vec k_3)\frac{\d\left[ k_1^3\langle\zeta_{k_1}^2\rangle'\right]}{\d\log k_1}\langle\zeta_{k_1}^2\rangle'\langle\zeta_{k_3}^2\rangle'\ .
\ee
Notice that since the short modes are still inside the horizon, their power spectrum is not yet scale invariant, so $\d\left[ k_1^3\langle\zeta_{k_1}^2\rangle'\right]/\d\log k_1$ is not slow-roll suppressed. Upon substitution of (\ref{eq:zetacl}), this is indeed equal to (\ref{eq:consistency_inside}), verifying the consistency condition for modes inside the horizon.

\subsection{Consistency condition for  operators with spatial derivatives}
  
Let us now consider the three-point function in the same regime of momenta as above for a derivative operator of the form
\be
\left\langle\frac{1}{a(\eta)^2}\d_i\zeta_{k_1}(\eta)\d_i\zeta_{k_2}(\eta)\zeta_{k_3}(\eta) \right\rangle  \ .
\ee
Since when we compute the three-point function we simply evolve the operators and not their spatial derivatives, the result can be trivially obtained from the one above in eq.~(\ref{eq:consistency_inside}) to be
\bea \nonumber
&&\left\langle\frac{1}{a(\eta)^2}\left(\d_i\zeta\right)_{k_1}(\eta)\left(\d_i\zeta\right)_{k_2}(\eta)\zeta_{k_3}(\eta)\right\rangle \simeq (2\pi)^3\delta^{(3)}(\vec k_1+\vec k_2+\vec k_3)\;\frac{k_1^2}{a(\eta)^2}\frac{1}{H} \d_t\langle\zeta_{k_1}^2\rangle'\;\langle\zeta_{k_3}^2\rangle'\\
&&\qquad=-\frac{ H^6}{8 \mpl^4 \epsilon ^2}\cdot\frac{\eta ^4\,k_1}{k_3^3}\ , \qquad\quad\ k_1\ll k_3\ .
\eea
This operator does not satisfy the consistency condition, that reads
\bea\label{eq:consistency_inside_derivative}
&&\left\langle\frac{1}{a(\eta)^2}\left(\d_i\zeta\right)_{k_1}(\eta)\left(\d_i\zeta\right)_{k_2}(\eta)\zeta_{k_3}(\eta)\right\rangle\simeq- (2\pi)^3\delta^{(3)}(\vec k_1+\vec k_2+\vec k_3)\frac{1}{a(\eta)^2}\frac{\d\left[ k_1^5\langle\zeta_{k_1}^2\rangle'\right]}{\d\log k_1}\langle\zeta_{k_1}^2\rangle'\langle\zeta_{k_3}^2\rangle'\ \nonumber \\ 
&&\qquad\qquad=-\frac{H^6 }{8   \mpl^4 \epsilon ^2}\frac{\eta ^2 \left(1+\eta ^2 k_1^2\right) }{k_1 k_3^3}\ .
\eea
The reason for this mismatch is that in the consistency condition we are rescaling all the momenta, including the ones representing the external derivatives. 

An operator that instead satisfies the consistency condition  (\ref{eq:consistency_inside_derivative}) is one in which the derivatives go together with factors of $e^{-\zeta_{k_3}}$. In the squeezed limit we have
\bea\label{eq:contact}
&&\left\langle\frac{1}{a(\eta)^2e^{2\zeta_{k_3}(\vec x,\eta)}}\left(\d_i\zeta\right)_{k_1}(\eta)\left(\d_i\zeta\right)_{k_2}(\eta)\zeta_{k_3}(\eta)\right\rangle=\\ \nonumber 
&&\left\langle\frac{1}{a(\eta)^2}\left(\d_i\zeta\right)_{k_1}(\eta)(\d_i\zeta)_{-k_1}(\eta)\zeta_{k_3\simeq0}(\eta)\right\rangle-2 \left\langle\frac{1}{a(\eta)^2}\left(\d_i\zeta\right)_{k_1}(\eta)\left(\d_i\zeta\right)_{-k_1}(\eta)\zeta_{k_3\simeq0}(\eta)\zeta_{-k_3\simeq0}(\eta)\right\rangle\ ,
\eea
as it can be readily verified.

We see that the consistency condition is satisfied by considering the sum of the operator we considered initially $\left(\d_i\zeta\right)_{k_1}\left(\d_i\zeta\right)_{k_2}\zeta_{k_3}$ plus a contact quartic operator of the form $-2\left(\d_i\zeta\right)_{k_1}\left(\d_i\zeta\right)_{k_2}\zeta_{k_3}{}^2 $. As we argued in the main text,  this additional contact operator comes automatically in the quartic Lagrangian, its presence being indeed guaranteed by the residual diff. invariance that we have in $\zeta$ gauge. The factor of $2$ apparent mismatch in the contact operator we insert in (\ref{eq:contact}) and the one we identify in the quartic Lagrangian in (\ref{eq:example}) takes into account the combinatorial factor that we have when we contract the operator with the external wavefunctions. This is the kind of combination of operators we consider in sec.~\ref{sec:cubic} when we use the consistency condition to show that some combination of diagrams do not lead to time dependence in the $\zeta$ correlators.

\subsection{Consistency condition for operators with time derivatives}\label{timederiv}

Here we want to show that time derivatives of operators, even when inside the horizon, will obey the consistency condition, when correlated with a long wavelength mode. In particular, we want to study a correlation function of the form $\langle \dot \zeta_{k_1}(\eta) \dot \zeta_{k_2}(\eta) \zeta_{k_3}(\eta) \rangle$ in the regime $k_3\ll k_1\approx k_2$, and the long mode has exited the horizon. 

As discussed in Sec.~\ref{sec:example1}, there is a contribution from contact terms that is essential for the consistency condition to be satisfied. For operators that involved spatial derivatives, we had to borrow terms from the quartic Hamiltonian. Here, we have a very similar situation.  For operators with time derivatives, these operators came naturally from $H_{4,3}$, i.e., the quartic Hamiltonian induced by the cubic Lagrangian. A more rigorous parallel between these cases is made at the end of this subsection. 

We will study an example in the Effective Field Theory of inflation where the speed of sound deviates from the speed of light, and the other background quantities, like $H$ and $\dot H$, are assumed to be constant, for effects of computing the tilt of the spectrum. The action was written in \eqref{eq:efttoyact} but let us write it before taking the decoupling limit:

\be
S=\int d^4x \sqrt{-g}\left[ \dot H \mpl^2 g^{00} +M^4(t) (\delta g^{00})^2 \right]\ .
\ee

Let us further assume that $M^4(t)$ has a linear dependence in time and we will concentrate only on the effects that are proportional to $\d_t(M^4)$. That is, we imagine that $M^4(t)$ varies on time scales that are slow with respect to $H^{-1}$, but fast with respect to $\epsilon H^{-1}$. Using the Stueckelberg procedure to recover gauge invariance, we perform a diffeomorphism $t \to t+\pi$ and consider the limit where the longitudinal mode decouples from the graviton. The action, up to cubic order, reads:
\be
S=\int dt d^3x a^3\; \left\{\frac{-\dot H \mpl^2}{c_s^2}\left[ \dot \pi^2 - c_s^2\left(\frac{\partial_i \pi}{a}\right)^2 \right]  + 4 \partial_t(M^4(t)) \pi \dot \pi^2 + 4 M^4(t)\dot \pi^3 -4 M^4(t) \dot \pi \left(\frac{\partial_i \pi}{a} \right)^2\right\}\ .
\ee
The speed of sound breaks the equivalent footing of time and space derivatives in the quadratic term, and is given by 
\be
c_s^2=\displaystyle{ -\dot H \mpl^2 \over 4 M^4(t) - \dot H \mpl^2}\ .
\ee

To write $\langle \dot\zeta_{k_1} \dot\zeta_{k_2} \zeta_{k_3} \rangle$ we first compute $\langle \dot\pi_{k_1} \dot\pi_{k_2} \pi_{k_3} \rangle$. In order to do it, we need the following operator equation, in Heisemberg picture:
\begin{align}\label{dotting}
\dot \pi(t) &= \partial_t(U_{int}^\dagger (t,-\infty_+\pi_I(t) U_{int}(t,-\infty_+))=\nonumber\\& i U_{int}^\dagger(t,-\infty_+)[H_{int}(t), \pi_I(t)] U_{int} (t,-\infty_+) +U_{int}^\dagger (t,-\infty_+)\dot\pi_I(t) U_{int} (t,-\infty_+)\ .
\end{align}
The quantum field $\pi$ can be written as $\pi_k(\eta) = a_{-k} \pi^{cl}(k,\eta) + a^\dagger_k \pi^{cl}(k,\eta)^*$, with the classical wavefunction given by:
\be
\pi^{cl}(k,\eta)=-{i \over 2 \sqrt{ \epsilon c_s k^3} \mpl}(1-i c_s k \eta) e^{i c_s k \eta} \ .
\ee
Then the three point function $\langle \dot \pi_{k_1}(\eta) \dot \pi_{k_2}(\eta) \pi_{k_3}(\eta) \rangle $ is given by:
\begin{align}
&\langle \dot \pi_{k_1}(\eta) \dot \pi_{k_2}(\eta) \pi_{k_3}(\eta)\rangle =i \int_{-\infty_+}^\eta d\tau \langle[H_3(\tau),\dot \pi_{k_1} (\eta)\dot \pi_{k_2} (\eta)\pi_{k_3} (\eta)] \rangle + \nonumber \\&+i \langle [H_3(\eta), \pi_{k_1}(\eta)] \dot \pi_{k_2}(\eta) \pi_{k_3}(\eta) \rangle + i \langle \dot \pi_{k_1}(\eta) [H_3(\eta), \pi_{k_2}(\eta)] \pi_{k_3}(\eta) \rangle\ .
\end{align}
The first term in the right hand side is the usual in-in expression, and the terms in the second line are the extra contact terms that come from using \eqref{dotting}. A straightforward computation of the three terms yields, in the squeezed limit:
\bea
&&i \int_{-\infty_+}^\eta d\tau \langle[H_3(\tau),\dot \pi_{k_1}(\eta) \dot \pi_{k_2}(\eta) \pi_{k_3}(\eta)] \rangle = (2\pi)^3 \delta^{(3)}\left(\sum k_i\right){1\over 8}\frac{c_s^4 \partial_t(M^4(t)) }{\mpl^6 \epsilon^3  } {(k_1 \eta)^4 \over k_1^3 k_3^3}\ ,\nonumber\\
&&i \langle [H_3(\eta), \pi_{k_1}(\eta)] \dot \pi_{k_2} (\eta)\pi_{k_3} (\eta)\rangle  =\\ \nonumber
&&\qquad\qquad =i \langle \dot \pi_{k_1} (\eta)[H_3(\eta), \pi_{k_2}(\eta)] \pi_{k_3}(\eta) \rangle = -(2\pi)^3 \delta^{(3)}\left(\sum k_i\right){1 \over 4}\frac{c_s^4 \partial_t(M^4(t)) }{ \mpl^6 \epsilon^3  } {(k_1 \eta)^4 \over k_1^3 k_3^3}\ .
\eea
So adding these terms will give us $\langle \dot \pi_{k_1}(\eta) \dot\pi_{k_2} (\eta)\pi_{k_3} (\eta)\rangle$. Now, we are interested in $\langle \dot\zeta_{k_1}(\eta) \dot\zeta_{k_2}(\eta)\zeta_{k_3} (\eta)\rangle$. But the $\zeta$ and $\pi$ fields are related through $\zeta = - H \pi+H\dot\pi\pi$~\cite{Cheung:2007sv}\footnote{There are additional quadratic corrections to this expression, but they will give corrections to the three point function that are subleading when at least one of the modes is outside of the horizon or that are slow roll suppressed.}, so we can write our desired correlator:
\bea
&&\langle \dot \zeta_{k_1}(\eta) \dot \zeta_{k_2}(\eta) \zeta_{k_3}(\eta) \rangle =\\ \nonumber
&&\qquad (2\pi)^3 \delta^{(3)}\left(k_1+k_2+k_3\right)\left[{3 \over 8} \frac{c_s^4 \partial_t(M^4(t)) H^3}{\mpl^6 \epsilon^3  } {(k_1 \eta)^4 \over k_1^3 k_3^3}+\frac{1}{H}\frac{\d\langle\dot\zeta_{k_1}^2\rangle'}{\d t}\langle\zeta_{k_3}^2\rangle'\right]\\ \nonumber
&&\qquad= - (2\pi)^3 \delta^{(3)}\left(k_1+k_2+k_3\right)\frac{c_s^3 H^4}{4\epsilon \mpl^2 }4\frac{(k_1\eta)^4}{k_1^3}\langle\zeta_{k_3}^2\rangle'=\\ \nonumber &&\qquad=-(2\pi)^3 \delta^{(3)}\left(k_1+k_2+k_3 \right)\frac{1}{k_1^3}\frac{d}{d \log k_1} \left(k_1^3\langle\dot \zeta_{k_1}^2 \rangle' \right) \langle\zeta_{k_3}^2 \rangle'\ , \quad k_3\ll k_1 \approx k_2\ ,
\eea
where we have used that
\bea
&&\langle \zeta_{k_1}(\eta) \zeta_{k_2}(\eta) \rangle  = (2\pi)^3 \delta^{(3)}(k_1+k_2)\frac{H^2(1+ c_s^2 k_1^2 \eta^2)}{4 c_s \epsilon \mpl^2 k_1^3}\ ,\\ \nonumber
&&\langle\dot \zeta_{k_1}(\eta) \dot\zeta_{k_2}(\eta) \rangle  = (2\pi)^3 \delta^{(3)}(k_1+k_2)\frac{c_s^3   H^4( k_1\eta)^4}{4  \epsilon \mpl^2\, k_1^3}\ .
\eea
Notice that the effect of the field redefinition is to remove the time derivatives associated to terms that do not depend explicitly on $k\eta$, such as $c_s$, so that the consistency condition works.
This concludes our check of the consistency condition for modes inside the horizon, with time derivative operators.

As a last remark, we now discuss the relation between the contact terms that contributed to $\langle \dot\pi\dot\pi\pi\rangle$, and the contact terms arising from the quartic Hamiltonian $H_{4,3}$, which is discussed in the main text. They are playing the exact same role: accounting for the action of the time derivative on $U_{int}$. The results of this subsection can be cast in a form that makes this connection more manifest. We use here the notation ``S, L" for short and long modes.

In the main text, we are computing a three point function of the following schematic form:
\be\label{relation}
\sum_a \int^\eta d\eta' \left \langle \left({\delta L_3 \over \delta {\cal D}^a \zeta^a}\right)_S(\eta') \zeta_L(\eta) \right \rangle {\cal D}^a G_\zeta(\eta',\eta) \sim \int^\eta d\eta' \left\langle i[H^{S,S,L,L}_{4,3}(\eta'),\zeta_L(\eta)] \zeta_L(\eta)\right\rangle+ \ldots\ ,
\ee
where $\ldots$ are contributions to the one-loop two point function coming from other diagrams. Now, we can recast the three point function $\langle \dot\pi\dot\pi\pi\rangle$ as:
\be
\langle \dot\pi_{S}(\eta) \dot\pi_{S}(\eta)\pi_{L}(\eta) \rangle=\left\langle\left(\delta L_3^{\pi \dot\pi^2}\over \delta \dot \pi\right)_S(\eta)  \; \pi_L(\eta) \right\rangle\ ,
\ee
and the contact term as:
\be
i \langle [H_3(\eta), \pi_{S}(\eta)] \dot \pi_{S}(\eta) \pi_{L}(\eta) \rangle \sim -\left \langle \left({\delta L_3\over \delta P}\right)_{S,L}(\eta)  \left({\delta L_3^{\pi \dot\pi^2} \over \delta \dot \pi}\right)_{S,L}(\eta) \right \rangle\ .
\ee
So we see that if the three point function involved the full Lagrangian, the contact term would be proportional to the squeezed quartic Hamiltonian, $\langle H_{4,3}^{S,S,L,L} \rangle$. As the one loop diagram involves a commutator instead of a tree level four point function, we need to insert the Green's function on the left hand side of \eqref{relation}, thus seeing how both three point functions are affected by contact terms coming from $H_{4,3}$.

\section{Local Anisotropic Universe\label{app:local_frame}}

We aim here to provide the change of coordinates that locally takes us from the metric written in standard $\zeta$ gauge to a form that is locally of the form of (\ref{eq:anisotropic_metric}). We need to work only at linear order in the long wavelength fluctuations $\zeta_L$, because in the loop we integrate over the short wavelength fluctuations $\zeta_S$. We start from the metric in ADM parametrization
\be
ds^2=-N^2 dt^2+\sum_{ij}\delta_{ij}a(t)^2e^{2\zeta}\left(dx^i+N^i dt\right)\left(dx^j+N^j dt\right)\ ,
\ee
where in this appendix we suspend the convention of summing over repeated indices. We can perform the following change of coordinates
\be
x^i=e^{\beta_{ij}(t)}\tilde x^j+C^i(t)\ ,
\ee
without introducing perturbations in the field that is driving inflation.
Since we can work at linear order in the long modes, we can use rotational invariance to consider a long mode with wavenumber only along the $\hat z$ direction,
%\be
%\zeta_L(\vec x,t)=\zeta_0(t) e^{i k z}\ ,
%\ee
\be
\zeta_L(\vec x, t)={\rm Re}\left[\tilde \zeta_0(t) \;e^{i k z}\right]\ ,\qquad {\rm Re}\left[\tilde \zeta_0\right]=\zeta_0\ .
\ee
It will be enough to take $\beta_{ij}=\beta(t) \delta_{i3}\delta_{j3}$. The only subtle point in this change of variables is that at linear order in the long modes, we have 
\be
\vec N_L=\left\{0,0,{\rm Re}\left[i\frac{\dot H}{H^2}\frac{1}{k}\dot{\tilde\zeta}_0 \, e^{i k\, e^\beta \tilde z}\right]\right\}+{\cal O}(k^2\tilde\zeta_0)\ ,
\ee
which does not have a nice behavior for $k\rightarrow 0$. We need therefore to enforce that our change of coordinates not only fixes to zero $N^i$ at one point, say the origin, $N^i_0=0$, but also it must set to zero $\d_i N^j$ at the origin, $(\d_i N^j)_0=0$. This will guarantee that neglected terms are suppressed in the limit $k\rightarrow 0$. 

Simple algebra shows that the solution is
\bea
&&\vec C=\int dt\left\{0,0,\frac{\dot H}{H^2}\frac{1}{k}{\rm Im}\left[\dot{\tilde\zeta}_0\right]\right\}\ , \\
&&\beta=\int dt\; \frac{\dot H}{H^2}\dot\zeta_0\ .
\eea
The metric then takes the form of (\ref{eq:anisotropic_metric}), with, in the new coordinates
\be
\tilde N^i_0=0\ , \qquad \left(\d_j \tilde N^i\right)_0=0\ , \qquad \tilde\zeta(\vec{\tilde x},t)=\zeta\left(\vec x(\vec{\tilde x},t),t\right)+\frac{2}{3} \int^tdt\;\frac{\dot H}{ H^2}\dot \zeta_0(t)\ .
\ee
Notice that the short mode fluctuations $\zeta_S$ transform as a scalar under this change of coordinates
\be
\tilde\zeta_S(\vec{\tilde x},t)=\zeta_S\left(\vec x(\vec{\tilde x},t),t\right)\ .
\ee
The same procedure can be clearly performed at non-linear level in $\zeta_L$ using a generic matrix $\beta_{ij}$, but this is not necessary for a one-loop calculation.

\end{appendix}

\end{document}